\documentclass[aps,pra,twocolumn,superscriptaddress,showkeys]{revtex4-2}
\usepackage{enumerate}
\usepackage{amsmath}
\usepackage{amssymb}
\begin{document}
\title{Hardy and Cabello Arguments in Spatial and Temporal Frauchiger–Renner Scenarios}
 \author{Ehsan Erfani Maharati}
\email{ehsan.erfani@um.ac.ir}
\affiliation{Department of Physics, Ferdowsi University of Mashhad, Mashhad, Iran}
\author{Ali Ahanj}
\email{a.ahanj@khayyam.ac.ir}
\affiliation{Department of Physics, Khayyam University, Mashhad, Iran}
\author{Mohsen Sarbishaei}
\email{sarbishaei@um.ac.ir}
\affiliation{Department of Physics, Ferdowsi University of Mashhad, Mashhad, Iran}
\begin{abstract}
We investigate Hardy- and Cabello-type logical structures within spatial
and temporal extensions of the Frauchiger--Renner (FR) framework,
embedding these constructions directly into the FR multi-observer architecture.
In the spatial multi-observer scenario, both Hardy and Cabello contradictions arise,
with the Cabello construction yielding the stronger violation,
\(\Delta_{\rm Cabello}^{\max}=0.1078\),
which exceeds the maximal Hardy probability
\(P_{H}^{\max}=\frac{5\sqrt{5}-11}{2}\approx 0.09017\).
We then develop a sequential temporal FR protocol based on coherent
multi-observer measurements performed on a single spin-$\tfrac12$ system.
In this temporal setting, the Hardy contradiction disappears identically
due to dynamical constraints imposed by sequential state updates,
whereas a finite Cabello-type violation survives,
\(\Delta_{\rm Cabello}^{\max}\approx0.0674\).
Our results establish a fundamental structural distinction between spatial
entanglement and temporal multi-observer correlations in FR-type logical scenarios,
and demonstrate that certain observer-independent description failures persist
even without spacelike separation.
\end{abstract}
\maketitle
\section{Introduction}
Thought experiments continue to play a central role in clarifying the conceptual foundations of quantum mechanics. Bell-type nonlocality \cite{Bell1964,CHSH1969}, temporal quantum correlations \cite{LeggettGarg1985}, and Wigner-type multi-observer scenarios \cite{Wigner1961,Frauchiger2018} each probe different tensions between quantum theory and classical notions of realism.

Bell’s theorem \cite{Bell1964} established that no local hidden-variable model can reproduce all quantum predictions. This incompatibility is captured quantitatively by the CHSH inequality \cite{CHSH1969}, whose maximal quantum violation is bounded by $2\sqrt{2}$, the Tsirelson limit \cite{Tsirelson1980}, and has been confirmed experimentally \cite{Aspect1982,Weihs1998,Hensen2015}. Hardy later provided an inequality-free proof of nonlocality \cite{Hardy1992,Hardy1993}, identifying configurations for which quantum theory predicts a nonzero joint probability that is strictly forbidden under local realism. Cabello extended this reasoning to a broader class of states, including mixed entangled states, under relaxed logical constraints and with finite success probability \cite{Cabello2002,Kunkri2006}.
 These logical approaches sharpen the distinction between quantum and classical correlations and have also proved useful in constraining post-quantum models \cite{Ahanj2019Cabello}.
 
 Temporal correlations exhibit related but structurally distinct departures from classical realism. The Leggett-Garg inequalities \cite{LeggettGarg1985} impose constraints on time-separated measurements under macrorealism and noninvasive measurability, assumptions that are incompatible with quantum dynamics. Hardy-type reasoning has been adapted to sequential measurements 
on a single system~\cite{Ahanj2007,Ahanj2010Hardy}, and 
subsequent work has developed temporal Bell inequalities and 
analyzed sequential measurement structures in 
detail~\cite{Aharonov2002,Palacios2010,Ahanj2007,Ahanj2010Hardy}.
While there are formal analogies with spatial nonlocality, the 
temporal case differs in that later measurements can alter the system's state, introducing an intrinsic asymmetry that is absent in spacelike-separated scenarios.

 The Wigner’s friend thought experiment \cite{Wigner1961} raises the question of whether measurement outcomes can be regarded as observer-independent. Frauchiger and Renner (FR) sharpened this issue in a multi-agent setting \cite{Frauchiger2018}, prompting extensive analysis. Brukner reformulated related setups as Bell-type tests \cite{Brukner2018}, leading to experimental investigations \cite{Proietti2019,Bong2020}. Although the original argument has been criticized on conceptual grounds \cite{Sudbery2017,Bub2021}, it has stimulated a broader structural reappraisal. Recent results reinterpret the Frauchiger–Renner scenario as a manifestation of contextuality rather than a direct inconsistency of universal unitarity. In particular, contextual agreement conditions resolve the apparent contradiction \cite{FedericoGrangier2023}, and it has been shown that any Wigner-type multi-agent paradox entails logical contextuality of Hardy form \cite{NurgalievaVilasini2025}.
 Although this result establishes a general connection between
Wigner-type scenarios and Hardy-form contextuality, it relies
on measurement compatibility and no-disturbance conditions
that are not satisfied in sequential temporal protocols;
we return to this point in Sec.~\ref{sec:discussion}.
Further refinements identify strong contextuality as the operative resource \cite{WalleghemBarbosaPuseyWeigert2026} and establish incompatibilities within generalized friendliness frameworks \cite{WalleghemCatani2025}, with complementary logical analyses clarifying the structural origin of the paradox \cite{BaltagSmets2024,Steane2025}. 
 
In this paper, we establish a unified connection between Hardy-, Cabello-, and Frauchiger--Renner-type logical structures. In Sec.~II, we embed Hardy- and Cabello-type constructions into a spatial FR protocol based on non-maximally entangled states. In Sec.~III, we introduce a temporal FR protocol involving sequential coherent measurements on a single spin-$\tfrac12$ system. Our analysis reveals a fundamental distinction between spatial and temporal multi-observer correlations: while the temporal protocol forbids a genuine Hardy contradiction, a finite Cabello-type violation nevertheless survives. Finally, in Sec.~IV, we summarize the main results and discuss their conceptual and operational implications for multi-observer quantum scenarios.
\section{Frauchiger--Renner (FR) Protocol with Hardy and Cabello Arguments}

The FR thought experiment \cite{Frauchiger2018} extends Wigner's friend paradox to a multi-agent scenario involving two nested laboratories. In each laboratory, an internal observer performs a projective measurement on a quantum system, while an external super-observer treats the entire laboratory as a coherent quantum object and may perform joint measurements on it. In the original protocol, Laboratory~1 contains a quantum coin prepared in superposition and measured by Alice, followed by a coherent measurement by Charlie on the combined system (coin + Alice). Conditioned on Alice’s outcome, a spin-$1/2$ particle is prepared and sent to Laboratory~2, where Bob measures the spin, and Debbie subsequently measures the joint system (spin + Bob).

Within this architecture, Frauchiger and Renner argue that three seemingly reasonable assumptions—(Q) the universal validity of quantum theory, (C) the consistency of observer-independent facts, and (S) the single-world character of outcomes—cannot be simultaneously maintained. Their incompatibility generates a logical tension that has motivated extensive debate on the foundations of quantum mechanics and the meaning of measurement.

Here we consider a reformulation of the FR protocol in which the original quantum coin is replaced by a non-maximally entangled bipartite source distributed to two spatially separated laboratories operated by Alice and Bob. Each laboratory is later probed coherently by a super-observer, Charlie and Debbie, respectively. This modification preserves the multi-agent structure of the FR scenario while providing a natural framework for embedding Hardy- and Cabello-type logical arguments directly into the FR architecture.

The source prepares a non-maximally entangled two-qubit state. In the computational basis
\(\{|+\rangle, |-\rangle\}\), the state is
\begin{equation}
|\psi(t_0)\rangle
=
\sum_{i=\pm}\lambda_i
| i\rangle_A |i\rangle_B ,
\label{eq:initial_state}
\end{equation}
where the Schmidt coefficients are
\(
\lambda_+=\cos(\theta/2)
\),
\(
\lambda_-=\sin(\theta/2)
\),
with the entanglement parameter satisfying
\(
0<\theta<\pi/2
\).
Here \(i=+\) corresponds to \(|+\rangle\) and \(i=-\) to \(|-\rangle\).
The state is distributed to Alice and Bob inside their respective laboratories. Including the ready states of the laboratories, the global state before the measurements is
\begin{equation}
|\Psi(t_1^-)\rangle
=\sum_{i=\pm}\lambda_i
|i\rangle_A |i\rangle_B
|A_0\rangle |B_0\rangle ,
\end{equation}
where
\(
|A_0\rangle
\)
and
\(
|B_0\rangle
\)
denote the ready states of Alice’s and Bob’s apparatuses (or memory registers).

Projective qubit measurements are implemented through real orthogonal basis transformations in the computational basis
\(
\{|+\rangle,|-\rangle\}
\).
We define the real orthogonal rotation matrix
\begin{equation}
\label{matrix}
R(\vartheta)=
\begin{pmatrix}
\cos\!\left(\frac{\vartheta}{2}\right) &
\sin\!\left(\frac{\vartheta}{2}\right)
\\
\sin\!\left(\frac{\vartheta}{2}\right) &
-\cos\!\left(\frac{\vartheta}{2}\right)
\end{pmatrix},
\end{equation}
which specifies a projective qubit measurement along a Bloch-sphere direction determined by the angle \(\vartheta\).

Following the local projective measurements performed by Alice and Bob at time \(t_1\), the corresponding local measurement bases are defined as
\begin{align}
|a_\alpha\rangle &=\sum_{i=\pm}
R_{i\alpha}(\theta_a)\,|i\rangle,
\qquad
\alpha\in\{+,-\},
\\
|b_\beta\rangle &= \sum_{j=\pm}
R_{j\beta}(\theta_b)\,|j\rangle,
\qquad
\beta\in\{+,-\}.
\end{align}
Thus, Alice and Bob perform projective spin measurements along directions
determined by the angles $\theta_a$ and $\theta_b$ relative to the
computational basis.
The measurement interactions correlate the spins with macroscopic memory states $|A_\alpha\rangle$ and $|B_\beta\rangle$ inside the laboratories.
 The measurement interactions correlate each spin with a macroscopic
memory state,
\(
|L_\alpha\rangle_1=|a_\alpha\rangle|A_\alpha\rangle
\)
and
\(
|L_\beta\rangle_2=|b_\beta\rangle|B_\beta\rangle
\).
\paragraph{Expansion in the $(A,B)$ basis.}
The post-measurement global state therefore takes the form
\begin{equation}
|\Psi(t_1^+)\rangle
=
\sum_{\alpha,\beta=\pm}
\Gamma_{\alpha\beta}^{AB}\,
|L_\alpha\rangle_1
|L_\beta\rangle_2 .
\end{equation}

All post-measurement amplitudes are compactly encoded in a single matrix relation. The coefficient matrix in the \((A,B)\) basis is defined as
\begin{equation}
\Gamma^{AB}
=
R(\theta_a)\,
\Lambda\,
R(\theta_b).
\end{equation}

Here,
\begin{equation}
\Lambda=
\begin{pmatrix}
\lambda_+ & 0 \\
0 & \lambda_-
\end{pmatrix},
\end{equation}
is the diagonal Schmidt matrix associated with the initial entangled state.

This representation generates all post-measurement amplitudes directly, without requiring the explicit trigonometric form of each coefficient.

\paragraph{Expansion in the $(C,B)$ basis.}
At time $t=t_2$, the super-observer Charlie performs a coherent measurement on Alice’s entire laboratory in the basis $\{|c_\gamma\rangle\}$, with $\gamma \in \{+,-\}$ where the basis states are related to the states $\{ | L_{\alpha}\rangle_1 \}$ via
\begin{equation}
     |c_{\gamma}\rangle = \sum_{\alpha=\pm}R_{\gamma \alpha}(\theta_c) |L_{\alpha}\rangle_{1}.
\end{equation}
The global wave function may equivalently be rewritten in the joint ((C,B)) basis as
\begin{equation}
|\Psi(t_2^+)\rangle
= \sum_{\gamma,\beta=\pm}
\Gamma_{\gamma\beta}^{CB}
|c_\gamma\rangle
|L_\beta\rangle_2 ,
\end{equation}
with the transformed coefficient matrix
\begin{equation}
\Gamma^{CB} = R(\theta_c)\Gamma^{AB}.
\end{equation}

At a later stage, specifically at time $t_3$, Debbie measures Bob's laboratory.  The measurement basis is $\{|d_\delta\rangle\}$, with $\delta \in \{ +, - \}$, where the basis states are related to the states $\{ | L_{\beta}\rangle_2 \}$ via
\begin{equation}
|d_\delta\rangle =\sum_{\beta=\pm} R_{\delta \beta}(\theta_d) |L_\beta\rangle_2.
\end{equation}
\paragraph{Expansion in the $(A,D)$ basis.}
After Debbie’s coherent measurement, the global state may be expressed as
\begin{equation}
|\Psi(t_3^+)\rangle
= \sum_{\alpha,\delta=\pm}
\Gamma_{\alpha\delta}^{AD}\, |L_\alpha\rangle
\,|d_\delta\rangle ,
\end{equation}
where the coefficient matrix is obtained from
\begin{equation}
\Gamma^{AD} = \Gamma^{AB}\, R(\theta_d),
\end{equation}
\paragraph{Expansion in the $(C,D)$ basis.}
Finally, the global state can then be written in the super–observer basis as
\begin{equation}
|\Psi(t_3^+)\rangle
=\sum_{\gamma,\delta=\pm}
\Gamma^{CD}_{\gamma\delta}
\,|c_\gamma\rangle |d_\delta\rangle ,
\end{equation}
where the transformed coefficient matrix is
\begin{equation}
\Gamma^{CD}
=R(\theta_c)\,
\Gamma^{AB}\,
R(\theta_d).
\end{equation}

\subsection{Hardy-Type Logical Structure}

Hardy’s argument \cite{Hardy1992,Hardy1993} can be formulated directly in terms of the coefficient matrices introduced above. In the present FR framework, the Hardy structure is naturally expressed through the joint expansions associated with Alice--Bob (A,B), Charlie--Bob (C,B), Alice--Debbie (A,D), and Charlie--Debbie (C,D), derived in the previous section.

The Hardy conditions read
\begin{align}
\text{(H1)}\quad & P(A=\alpha,B=\beta)=|\Gamma^{AB}_{\alpha\beta}|^2=0, \\
\text{(H2)}\quad & P(A=\bar\alpha,D=\delta)=|\Gamma^{AD}_{\bar\alpha\delta}|^2=0, \\
\text{(H3)}\quad & P(C=\gamma,B=\bar\beta)=|\Gamma^{CB}_{\gamma\bar\beta}|^2=0, \\
\text{(H4)}\quad & P(C=\gamma,D=\delta)=|\Gamma^{CD}_{\gamma\delta}|^2 \geq 0 .
\end{align}
Let \(\alpha,\beta,\gamma,\delta\in\{+,-\}\) and denote by \(\bar\alpha\) and \(\bar\beta\) the opposite outcomes. The coefficient matrices are those defined in Sec.~II.

From these conditions, one can derive an inconsistency with any single-world probabilistic description that assigns jointly well-defined outcomes to all measurement contexts. More precisely, assuming the validity of (H1)–(H3) together with the existence of a global joint probability distribution over all observables consistent with noncontextual outcome assignments, one obtains an upper constraint on $P(C=\gamma, D=\delta)$ imposed by the logical structure of the model. However, quantum mechanics predicts a strictly positive value for this probability, thereby producing a Hardy-type contradiction within the Frauchiger–Renner framework.

The maximum quantum mechanical value of \(P(C=\gamma, D=\delta)\) compatible with (H1)–(H3) is \cite{Hardy1993}
\begin{equation}
P_{\text{max}}(C=\gamma, D=\delta) = \frac{5\sqrt{5}-11}{2} \approx 0.09017 .
\end{equation}
The constraints (H1)–(H3) are incompatible with any framework that admits a single, context-independent joint probability distribution over all measurement outcomes associated with the different experimental settings. Within such a framework, the imposed zero-probability relations enforce a vanishing contribution for the event $(C=\gamma, D=\delta)$. In contrast, quantum mechanics predicts a strictly positive value for this probability,
\(
P(C=\gamma, D=\delta) > 0,
\)
thereby giving rise to a Hardy-type contradiction within the Frauchiger–Renner structure.


We emphasize that the Hardy event arises only for non–maximally entangled states ($0<\theta<\pi/2$) and disappears in the maximally entangled limit $\theta=\pi/2$. No additional hidden-variable assumptions are invoked; the derivation remains entirely within the standard conditional-probabilistic structure of quantum theory. The detailed calculation is provided in Appendix~\ref{app:hardy_calc}.
\subsection{Cabello-Type Logical Contradiction}

Cabello's argument \cite{Cabello2002} extends the Hardy-type structure by relaxing one of the strict zero-probability constraints. In the present Frauchiger--Renner (FR) framework, condition (H1) is replaced by the weaker requirement
\begin{equation}
P(A=\alpha,B=\beta)\neq 0,
\end{equation}
while conditions (H2), (H3), and the non-negativity constraint (H4) are retained. The associated logical tension is quantified by
\begin{equation}
\Delta_{\rm Cabello} := P(C=\gamma,D=\delta) - P(A=\alpha,B=\beta),
\label{eq:Delta_Cabello}
\end{equation}
whose maximization under the FR measurement constraints yields
\begin{equation}
\Delta_{\rm Cabello}^{\rm max} = 0.1078.
\label{eq:Cabello_max_value}
\end{equation}
This maximum is obtained for suitable choices of the measurement angles (see Appendix~\ref{app:cabello}). Notably, it exceeds the maximal Hardy probability $0.09017$, indicating that relaxing a zero-probability condition leads to a stronger quantitative separation between quantum predictions and single-world descriptions.

Within the FR protocol, the logical structure underlying the Cabello construction can be summarized as follows:

\begin{enumerate}[(i)]
\item From (H3): $P(C=\gamma, B=\bar\beta)=0$, implying a deterministic inference from Charlie’s outcome to Bob’s result.
\item Condition (H1) is relaxed, so $P(A=\alpha,B=\beta)\neq 0$, allowing compatible occurrences of Alice’s and Bob’s outcomes.
\item From (H2): $P(A=\bar\alpha, D=\delta)=0$, constraining the correlation between Alice’s and Debbie’s outcomes.
\end{enumerate}
These relations constrain the possible joint assignments across observers and lead to a nontrivial bound on $P(C=\gamma,D=\delta)$ when outcomes are assumed to be consistently defined across the protocol. Quantum mechanics, however, predicts a violation of this bound, quantified by $\Delta_{\rm Cabello}$.

In this sense, the Cabello construction provides a stronger quantitative separation than the Hardy case within the FR framework.

Embedding the Hardy and Cabello constructions into the FR framework provides a unified probabilistic formulation of multi-observer quantum contradictions. The resulting logical tensions follow solely from the joint assumptions (Q) universal validity of quantum theory, (C) consistency of observer records, and (S) single-world definiteness, without requiring counterfactual reasoning. This demonstrates that the origin of the contradiction is structural rather than inferential: it reflects the impossibility of consistently assigning a single joint probability distribution across different observer perspectives under the constraints (Q), (C), and (S). Consequently, the FR scenario admits a direct operational interpretation in terms of experimentally accessible probabilities while preserving a logical structure that generalizes Hardy- and Cabello-type nonclassical correlations.
\section{Temporal Extension: Successive Measurements in Multi-Observer Framework}
In this section we introduce a temporal extension of the FR
framework in which spatial entanglement between distinct subsystems is replaced
by temporal correlations generated through successive measurements on a single
quantum system. The goal is to investigate whether Hardy- and Cabello-type logical
constraints can be consistently implemented when the multi-observer structure of
the FR protocol is embedded into a sequential measurement setting.
\subsection{Motivation and Conceptual Framework}
Temporal correlations constitute a well-established source of nonclassical
behavior in quantum mechanics. Unlike spatial Bell scenarios, where correlations
arise from entanglement shared between spacelike-separated systems, temporal
correlations emerge from the interplay between measurement back-action and
subsequent unitary evolution. This distinction becomes especially significant in
multi-observer contexts, where inference chains rely on assumptions about
observer-independent facts across different measurement stages.

Previous works have shown that temporal analogues of Bell-type inequalities can
be violated by quantum mechanics \cite{LeggettGarg1985,Ahanj2010Hardy}.
However, the compatibility of temporal correlations with multi-observer reasoning
of the Frauchiger--Renner type has not been systematically analyzed. In
particular, it is not a priori clear whether logical nonlocality arguments that
rely on strict zero-probability constraints—such as Hardy’s argument—can survive
the inherently sequential nature of temporal measurements.
While Hardy-type arguments for successive spin measurements have been studied
previously~\cite{Ahanj2010Hardy}, the present analysis differs in two
essential respects: we embed the sequential measurement protocol within the
full Frauchiger--Renner multi-observer architecture including coherent super-observer measurements on entire laboratories, and we introduce the Cabello generalization into this temporal setting for the first time.

\subsection{Experimental Setup and Fundamental Assumptions}
The temporal FR protocol employs sequential measurements on a single spin-$\tfrac12$ system. Alice (Laboratory~1) first performs a projective measurement along \(\mathbf a\) without outcome selection. The same spin is then measured by Bob (Laboratory~2) along \(\mathbf b\) at a later time \(t_2 > t_1\). The two measurements are distinguished purely by their temporal order, not by spatial separation.

The measurement outcomes of Alice and Bob are stored in orthogonal memory states associated with their respective laboratories. Charlie and Debbie subsequently act as super-observers associated with Laboratories~1 and~2. Each super-observer is assumed capable of performing coherent measurements on the composite system consisting of the measured spin together with the corresponding observer’s memory register.

Operationally, the protocol may be viewed as a sequential Stern--Gerlach architecture in which a single spin passes successively through two measurement stages. The first stage correlates the spin with Alice’s memory states, while the second stage correlates the same spin with Bob’s memory states. The super-observer measurements correspond to coherent basis rotations and joint measurements performed on the enlarged laboratory Hilbert spaces.

The analysis assumes ideal projective measurements with unit efficiency and strictly unitary evolution between measurement stages. No decoherence is assumed during transmission or storage of the quantum information. Quantum theory is taken to apply universally, including to measurement devices and internal observers themselves. Alice and Bob are therefore treated as quantum systems that can, in principle, be measured coherently by Charlie and Debbie.

The framework further assumes single-world consistency and the possibility of assigning observer-independent measurement records across different stages of the protocol. At the same time, the temporal ordering of measurements introduces unavoidable state-update effects absent in spacelike-separated Bell scenarios. Consequently, the resulting logical structure is governed not only by measurement correlations but also by the dynamical constraints imposed by sequential quantum evolution.

Taken together, these assumptions define the operational setting within which Hardy- and Cabello-type logical structures are analyzed in the temporal multi-observer scenario.
\subsection{Temporal FR Protocol: Detailed Implementation}

The temporal FR protocol involves two internal observers, Alice and Bob, who perform sequential projective measurements on a single spin-$\tfrac12$ system at times $t_1$ and $t_2$, respectively. Two external observers, Charlie and Debbie, subsequently act as super-observers and perform coherent measurements on the composite laboratories formed by the internal observers and the measured particle at times $t_3$ and $t_4$. The temporal ordering is
\[
t_0 < t_1 < t_2 < t_3 < t_4 .
\]

\paragraph*{Stage 1 ($t=t_0$): Initial state preparation.}
A single spin-$\tfrac12$ particle is prepared in the state $\lvert +\rangle$, corresponding to spin-up along the $z$ axis. Alice’s apparatus is initialized in the ready state $\lvert A_{0}\rangle$, so that the total state reads
\begin{equation}
\lvert \Psi(t_0^+) \rangle
= \lvert A_{0}\rangle \otimes \lvert +\rangle .
\end{equation}

\paragraph*{Stage 2 (\(t=t_1\)): Alice’s projective interaction.}

The spin is coupled to Laboratory~1 via an ideal projective interaction along direction $\mathbf a$, specified by the polar angle $\theta_a$ (with azimuth $\phi_a=0$ without loss of generality). The measurement basis is
\begin{align}
 |a_{\alpha}\rangle = \sum_{i=\pm}R_{i\alpha}(\theta_a) |\alpha\rangle,\quad \alpha\in\{+,-\} .
\end{align}

The interaction entangles the spin with orthogonal pointer states of Alice’s laboratory. The post-interaction state becomes
\begin{equation}
\lvert \Psi(t_1^+) \rangle
= \sum_{\alpha=\pm} a_\alpha  \lvert L_\alpha\rangle_{1} ,
\end{equation}
where $|L_\alpha\rangle_1
=
|a_\alpha\rangle |A_\alpha\rangle ,$
with amplitudes 
\begin{equation}
a_+
=
\cos\frac{\theta_a}{2},
\qquad
a_-
=
\sin\frac{\theta_a}{2}.
\end{equation}
\paragraph*{Stage 3 ($t=t_2$): Bob's projective measurement.}
Alice sends her spin to Bob. Upon receiving the particle, Bob measures it along direction $\mathbf b$ with polar angle $\theta_b$. The basis states of observer $B$ can be expressed in terms of the basis states of observer $A$ via the rotation matrix $R(\theta_{ba})$:
\begin{equation}
|b_\beta\rangle = \sum_{\alpha=\pm} R_{\beta\alpha}(\theta_{ba}) \, |a_\alpha\rangle, \qquad \beta \in \{+,-\},
\end{equation}
where 
\(
R(\theta_{ba}) = R(\theta_b) R^T(\theta_a),
\)
using the orthogonal rotation matrix \(R(\theta)\) defined in Eq.~(\ref{matrix}) (which satisfies \(R^{-1}(\theta)=R^T(\theta)=R(-\theta)\)). For the special case where all rotations are about the same axis (e.g., \(\phi=0\)), one obtains \(R(\theta_{ba}) = R(\theta_b - \theta_a)\). Explicitly,
\begin{align}
|b_+\rangle &= \cos\frac{\theta_{ba}}{2} \, |a_+\rangle - \sin\frac{\theta_{ba}}{2} \, |a_-\rangle, \\
|b_-\rangle &= \sin\frac{\theta_{ba}}{2} \, |a_+\rangle + \cos\frac{\theta_{ba}}{2} \, |a_-\rangle.
\end{align}
The measurement updates Bob's memory state according to
\begin{equation}
|b_\beta\rangle |B_0\rangle \;\longrightarrow\; |b_\beta\rangle |B_\beta\rangle,
\end{equation}
where $\{|B_\beta\rangle\}$ are orthogonal memory states. Immediately after Bob's interaction, the global state becomes
\begin{equation}
|\Psi(t_2^+)\rangle = \sum_{\alpha,\beta} b_{\alpha\beta} \, |L_\alpha\rangle_1  |L_\beta\rangle_2,
\end{equation}
with the nonvanishing amplitudes
\begin{align*}
b_{++} &= +\cos\frac{\theta_a}{2}\cos\frac{\theta_{ba}}{2}, &
b_{+-} &= +\cos\frac{\theta_a}{2}\sin\frac{\theta_{ba}}{2}, \\[4pt]
b_{-+} &= +\sin\frac{\theta_a}{2}\sin\frac{\theta_{ba}}{2}, &
b_{--} &= -\sin\frac{\theta_a}{2}\cos\frac{\theta_{ba}}{2}.
\end{align*}
Accordingly, the joint probability for Alice and Bob's outcomes is given by
\begin{equation}
P(A=\alpha, B=\beta) = |b_{\alpha\beta}|^2.
\end{equation}
\paragraph*{Stage 4 ($t=t_3$): Charlie’s coherent probe of Alice’s laboratory.}

Charlie performs a coherent measurement on Alice’s entire laboratory, including both the spin degree of freedom and Alice’s memory register.
\begin{equation}
\label{umatrix}
\lvert c_\gamma\rangle
=
\sum_{\alpha=\pm} R_{\alpha\gamma}(\theta_c)
\lvert L_\alpha\rangle_1 ,
\end{equation}
with $\gamma\in\{+,-\}$. Starting from the pre-measurement state $\lvert \Psi(t_3^-) \rangle
= \lvert \Psi(t_2^+)\rangle $, Charlie’s coherent interaction yields
\begin{equation}
\lvert \Psi(t_3^+) \rangle
=
\sum_{\gamma=\pm}\lvert c_\gamma\rangle
\sum_{\beta=\pm} \sum_{\alpha=\pm}g_{\gamma\beta}
\lvert L_\beta\rangle_2,
\end{equation}
and the total joint amplitude is obtained by summing over Alice’s branches,
\begin{equation}
g_{\gamma\beta} = \sum_{\alpha=\pm} R_{\gamma\alpha}(\theta_c) b_{\alpha \beta}.
\end{equation}
Accordingly, the joint probability for Charlie obtaining outcome $\gamma$ and Bob obtaining outcome $\beta$ is given by
\begin{equation}
P(C=\gamma, B=\beta) = \left| g_{\gamma\beta} \right|^2 .
\end{equation}
\paragraph*{Stage 5 ($t=t_4$): Debbie’s coherent probe of Bob’s laboratory.}

Debbie now performs a coherent measurement on Bob’s entire laboratory, including
the spin and Bob’s memory record. Her cat basis is
\begin{equation}
\lvert d_\delta\rangle
=\sum_{\beta=\pm} R_{\beta\delta}(\theta_d)\,
\lvert L_\beta\rangle_{2},
\end{equation}
with $\delta\in\{+,-\}$ and
$\lvert \Psi(t_4^-)\rangle=\lvert \Psi(t_3^+)\rangle$.

\noindent\textit{Alice--Debbie statistics.}—
To obtain the joint statistics for Alice and Debbie, we express the composite
state immediately before Debbie’s interaction in Alice’s measurement basis and
Debbie’s cat basis,
\begin{equation}
\lvert \Psi(t_4^-) \rangle
=
\sum_{\alpha=\pm}
\lvert L_\alpha\rangle_1 
\sum_{\delta=\pm}
e_{\alpha\delta}\, \lvert d_\delta\rangle .
\end{equation}
The branch-resolved amplitudes are obtained by projecting Bob’s laboratory onto
Debbie’s measurement basis,
\begin{equation}
e_{\alpha\delta}
=
\sum_{\beta=\pm}
b_{\alpha\beta}\, R_{\beta\delta}(\theta_d),
\end{equation}
where \(b_{\alpha\beta}\) is the amplitude from Alice $\rightarrow $ Bob measurement in Stage~2. Accordingly, the joint probability for Alice obtaining outcome \(\alpha\) and Debbie
obtaining outcome \(\delta\) is
\begin{equation}
P(A=\alpha, D=\delta) = \left| e_{\alpha\delta} \right|^2 .
\end{equation}

\noindent\textit{Charlie--Debbie statistics.}—
By inserting the inverse Bob-to-Debbie transformation into the post-Charlie
state, the global state after Debbie’s coherent probe can be written as
\begin{equation}
\lvert \Psi(t_4^+) \rangle
=
\sum_{\gamma=\pm}\sum_{\delta=\pm}
f_{\gamma\delta}
\,\lvert c_\gamma\rangle \lvert d_\delta\rangle ,
\end{equation}
where the branch-dependent amplitudes are defined as
\begin{equation}
f_{\gamma\delta}
\equiv \sum_{\alpha=\pm}\sum_{\beta=\pm}
R_{\gamma\alpha}(\theta_c)\, b_{\alpha \beta} R_{\beta\delta}(\theta_d).
\end{equation}
Summing coherently over Alice’s branches, the joint super-observer probabilities
for Charlie and Debbie are therefore given by
\begin{equation}
P(C=\gamma, D=\delta)
=
\left|
f_{\gamma\delta}\right|^2 .
\end{equation}

The resulting protocol implements a sequential, temporal analogue of the
Frauchiger--Renner scenario, in which automatic state-dependent routing and nested
super-observer measurements generate distinct joint probability distributions.
This architecture provides a natural setting for testing Hardy- and
Cabello-type temporal correlations without requiring simultaneous entanglement
between spatially separated laboratories.
\section{Hardy- and Cabello-Type Logic in the Temporal FR Protocol}

We now analyze logical nonclassicality in the temporal FR protocol using Hardy- and Cabello-type structures adapted to sequential measurements. In contrast to spatial Bell scenarios, the temporal architecture involves explicit quantum state update after each observer’s intervention. As a result, the logical constraints linking different observers are dynamically correlated rather than kinematically independent.

\subsection{Hardy-Type Constraints}
The temporal Hardy structure is specified by the conditions
\begin{align}
P(A=\alpha,B=\beta)=|b_{\alpha\beta}|^2 &= 0, \label{H1}\\
P(A=\bar\alpha,D=\delta)=|e_{\bar\alpha\delta}|^2 &= 0, \label{H2}\\
P(C=\gamma,B=\bar\beta)=|g_{\gamma\bar\beta}|^2 &= 0, \label{H3}
\end{align}
together with the requirement
\begin{equation}
P(C=\gamma,D=\delta)=|f_{\gamma\delta}|^2 > 0. \label{H4}
\end{equation}

In a classical single-history model these constraints are mutually inconsistent: if the first three probabilities vanish, the last one must also vanish.

In the temporal FR protocol, the joint probabilities are generated by sequential unitary embeddings. Using the amplitudes defined in the previous section, Eqs.~(\ref{H1})--(\ref{H3}) impose functional relations among the measurement angles. The detailed derivation is presented in Appendix~\ref{app:hardy_temp}.

The key result is that under these constraints one obtains
\begin{equation}
P(C=\gamma,D=\delta) = 0.
\end{equation}
Hence, unlike the spatial Hardy paradox, the temporal FR protocol does not admit a genuine Hardy contradiction: the same dynamical relations that enforce the zero-probability events also force the supposedly nonzero event to vanish. The temporal update structure therefore suppresses Hardy-type violations.
\subsection{Cabello-Type Generalization}

The Cabello logic relaxes the first Hardy constraint and considers instead the difference
\begin{equation}
\Delta_{\rm Cabello} = P(C=\gamma,D=\delta) - P(A=\alpha,B=\beta),
\label{eq:Deltadef}
\end{equation}
subject to the remaining zero-probability conditions
\begin{equation}
P(A=\bar\alpha,D=\delta) = 0, \qquad P(C=\gamma,B=\bar\beta) = 0.
\label{eq:Cabello_zeros}
\end{equation}
Classically one must have $\Delta_{\rm Cabello} \le 0$, whereas quantum theory may allow $\Delta_{\rm Cabello} > 0$.

In the temporal FR protocol, using the amplitudes defined in Sec.~III, the constraints in Eq.~(\ref{eq:Cabello_zeros}) impose functional relations among the measurement angles. The detailed derivation is presented in Appendix~\ref{app:cabello_temp}. The key result is that the maximum achievable value of $\Delta_{\rm Cabello}$ is
\begin{equation}
\Delta_{\rm Cabello}^{\max} \approx 0.0674.
\label{eq:Cabello_max}
\end{equation}
This positive value, although smaller than the spatial Cabello maximum ($0.1078$), demonstrates that a Cabello-type logical contradiction persists in the temporal FR protocol despite the suppression of the Hardy-type violation. The temporal update structure therefore allows a weaker but still nonclassical logical tension, where the difference of two probabilities becomes positive while two other joint probabilities vanish.

\section{Discussion and Conclusions}
\label{sec:discussion}
We have examined Hardy- and Cabello-type logical constructions within both spatial and temporal implementations of the Frauchiger–Renner (FR) framework. Our analysis reveals a clear structural distinction between the two settings: spatial multi-observer scenarios rely on the kinematic independence of spacelike-separated subsystems, whereas temporal FR protocols are governed by sequential unitary evolution and explicit quantum state updates. This difference directly shapes the strength and, in some cases, the very existence of logical contradictions in multi-observer quantum scenarios.
\subsection{Spatial FR scenario}

In the spatial FR setting, realized with a non-maximally entangled state shared between laboratories, the Hardy argument yields a genuine nonlocal contradiction. The maximum probability for the Hardy event is
\begin{equation}
P_{\rm Hardy}^{\max} = \frac{5\sqrt{5}-11}{2} \approx 0.09017.
\end{equation}
When the first Hardy condition is relaxed, the Cabello construction provides a stronger separation, quantified by the difference
\begin{equation}
\Delta_{\rm Cabello}^{\rm spatial} = P(C=\gamma,D=\delta) - P(A=\alpha,B=\beta),
\end{equation}
which reaches a maximum value of
\begin{equation}
\Delta_{\rm Cabello}^{\rm spatial, max} = 0.1078.
\end{equation}
Thus, within spatial multi-observer architectures, Cabello's logic offers a quantitatively more powerful witness of the incompatibility between quantum predictions and any single-world, observer-independent account.

\subsection{Temporal FR scenario}
In contrast, we have developed a detailed temporal extension of the FR protocol, in which a single spin-\(1/2\) system is measured sequentially by Alice, Bob, Charlie, and Debbie, with coherent super-observer measurements performed on entire laboratories. To the best of our knowledge, the comparative analysis of Hardy- and Cabello-type logical structures within a sequential Frauchiger--Renner framework has not been systematically explored.

When the temporal Hardy conditions are enforced, we find that the dynamical constraints impose
\begin{equation}
\theta_{ba} = \pi, \quad \theta_c = \pi, \quad \theta_d = \pi,
\end{equation}
which force the remaining event probability to vanish:
\begin{equation}
P(C=+,D=+) = 0.
\end{equation}
Hence, no genuine Hardy-type contradiction arises in the temporal protocol. The mechanism that enforces the zero-probability events simultaneously removes the possibility of a nonzero Hardy event, demonstrating a fundamental difference from the spatial case.

By contrast, the Cabello structure remains operative in the temporal protocol even though the Hardy violation is suppressed. Imposing only the two zero-probability conditions
\begin{equation}
P(A=-,D=+) = 0, \qquad P(C=+,B=-) = 0,
\end{equation}
and optimizing the remaining free parameters yields a strictly positive maximum for the difference:
\begin{equation}
\Delta_{\rm Cabello}^{\rm temporal, max} \approx 0.0674.
\end{equation}
This value, while smaller than the spatial Cabello maximum, demonstrates that even without spatial entanglement or genuine nonlocality, the temporal multi-observer architecture cannot be embedded into a single consistent classical narrative. The temporal protocol thus exhibits a form of \emph{temporal contextuality} arising from coherent inclusion of successive observers rather than from spacelike separation.

Our temporal no-Hardy result appears to stand in tension with
the general claim of Ref.~\cite{NurgalievaVilasini2025} that
every Wigner-type multi-agent paradox entails Hardy-form logical
contextuality.
The resolution lies in the structural assumptions underlying
that theorem.
The Nurgalieva--Vilasini result applies to measurement scenarios
satisfying \emph{local friendliness} and \emph{measurement
compatibility} conditions: specifically, it assumes that the
joint distributions associated with different agent perspectives
can be embedded in a common probability space, with
no-disturbance conditions holding across measurement contexts.
In our temporal protocol, the sequential measurement back-action
explicitly violates no-disturbance: Bob's projective measurement
at time $t_2$ physically disturbs the spin state that Alice
measured at $t_1$, breaking the compatibility structure that
underpins the Nurgalieva--Vilasini theorem.
Formally, the post-measurement state
$|\Psi(t_2^+)\rangle$ depends irreducibly on the outcome of
Alice's earlier interaction, so the joint distributions
$P(A,B)$, $P(C,B)$, $P(A,D)$, and $P(C,D)$ do not arise
from a single non-disturbing measurement scheme.
The temporal FR scenario therefore falls outside the scope of
that result---not by contradiction, but by a failure of its
preconditions.
Conversely, the survival of a Cabello-type violation in our
protocol is consistent with contextuality results that do not
require measurement compatibility, since the Cabello witness
imposes only two strict zero-probability conditions rather than
the full Hardy structure.
\paragraph{Contextuality-theoretic interpretation.}
The present results can be rephrased within the resource-theoretic
framework of quantum contextuality.
In the spatial FR scenario, both Hardy and Cabello witnesses
detect \emph{strong contextuality} of the joint agent probability
distributions: the Hardy witness certifies that no
noncontextual hidden-variable model can reproduce all four
joint distributions simultaneously, while the Cabello witness
provides a strictly positive quantitative gap even after
relaxing one zero-probability condition.
These findings are consistent with the analysis of
Refs.~\cite{WalleghemBarbosaPuseyWeigert2026,
NurgalievaVilasini2025}, which identify strong contextuality
as the operative resource in Wigner-type multi-agent paradoxes.

In the temporal protocol, the vanishing of the Hardy witness
indicates that \emph{strong contextuality}---in the sense
requiring strict zero-probability events in all but one
context---is destroyed by sequential measurement back-action.
The dynamical state update after each observer's intervention
prevents the existence of a consistent global assignment of
outcomes across all four joint distributions simultaneously,
but it does so in a way that also eliminates the Hardy event
itself.
Nevertheless, the survival of the Cabello witness demonstrates
that a \emph{weaker form of contextuality}---involving only two
strict zero-probability events and a positive probability
difference---persists under sequential evolution.
This hierarchy between Hardy-type and Cabello-type contextuality
witnesses maps directly onto the structural distinction between
spatial and temporal multi-observer correlations: spatial
entanglement supports the stronger Hardy form, whereas temporal
correlations generated by sequential disturbance support only
the weaker Cabello form.
The temporal FR protocol therefore provides a concrete example
in which the strength of contextuality is degraded, but not
eliminated, by the inherent asymmetry of sequential quantum
measurements.

\subsection{Experimental implications}

From an experimental standpoint, the temporal FR protocol can in principle be simulated on current few-qubit platforms. Sequential spin measurements and controlled spin–memory couplings are routinely implemented in trapped-ion and nitrogen-vacancy centers. Internal observers (Alice and Bob) can be represented by auxiliary qubits, while super-observer measurements (Charlie and Debbie) correspond to multi-qubit entangling operations and global basis rotations. Although macroscopic laboratory superpositions are not presently feasible, coherent few-qubit implementations sufficient to realize the logical structure of the protocol are compatible with existing quantum technologies. Our analysis provides explicit angle parameters that can be directly used in such experiments.
\paragraph{Minimal circuit implementation.}
The temporal FR protocol can be mapped onto a minimal
quantum register consisting of three qubits: one \emph{system
qubit} $S$ representing the spin-$\tfrac12$ particle, one
\emph{memory qubit} $M_A$ encoding Alice's record
$\{|A_\alpha\rangle\}$, and one \emph{memory qubit} $M_B$
encoding Bob's record $\{|B_\beta\rangle\}$.
Two additional ancilla qubits $M_C$ and $M_D$ are required
if Charlie's and Debbie's coherent measurements are to be
implemented as explicit entangling operations rather than
classical post-processing of the memory registers.
The gate sequence implementing the five-stage protocol is as
follows:
\begin{enumerate}[(1)]
\item \textbf{State preparation.}
      Initialize $S$ in $|{+}\rangle$ and all memory qubits
      in $|0\rangle$.
\item \textbf{Alice's interaction.}
      Apply a controlled rotation
      $\mathrm{CR}_y(\theta_a)$ between $S$ (control) and
      $M_A$ (target), entangling the spin with Alice's memory
      at angle $\theta_a$.
\item \textbf{Spin transport to Bob.}
      Apply a single-qubit rotation $R_y(\theta_{ba})$ on $S$
      to implement the relative measurement angle between
      Alice and Bob.
\item \textbf{Bob's interaction.}
      Apply a controlled rotation $\mathrm{CR}_y(\theta_{ba})$
      between $S$ (control) and $M_B$ (target), correlating
      the spin with Bob's memory.
\item \textbf{Charlie's coherent measurement.}
      Apply a two-qubit basis rotation
      $U_C(\theta_c)$ on the composite register
      $(S, M_A)$, implementing the super-observer basis
      $\{|c_\gamma\rangle\}$ defined in Eq.~\eqref{umatrix}.
\item \textbf{Debbie's coherent measurement.}
      Apply a two-qubit basis rotation $U_D(\theta_d)$
      on the composite register $(S, M_B)$, implementing
      the super-observer basis $\{|d_\delta\rangle\}$.
\item \textbf{Measurement.}
      Measure all qubits in the computational basis and
      record the outcomes for statistical estimation of
      $P(A,B)$, $P(C,B)$, $P(A,D)$, and $P(C,D)$.
\end{enumerate}
The super-observer operations in steps~(5) and~(6) are
two-qubit entangling rotations; single-qubit and two-qubit
controlled gates of this type are routinely implemented in
trapped-ion~\cite{Bruzewicz2019} and superconducting-qubit~\cite{Krantz2019} platforms with
fidelities exceeding $99\%$.
The total circuit depth is five layers (two single-qubit
rotations and three entangling gates), well within the
coherence limits of current hardware.

\paragraph{Noise robustness.}
The temporal Cabello gap $\Delta_{\rm Cabello}^{\max}\approx
0.0674$ is modest in magnitude, and its experimental
observability therefore depends on the achievable gate and
readout fidelity.
With typical state-preparation-and-measurement (SPAM) error
rates of $\epsilon\sim 1\%$ on current trapped-ion platforms,
the effective reduction of the observed gap is at most
$O(4\epsilon)\sim 4\%$ of the gap value, since each of the
four joint probabilities entering
$\Delta_{\rm Cabello}$ carries an independent SPAM
contribution~\cite{Bruzewicz2019}.
The corrected gap $\Delta_{\rm Cabello}^{\rm obs}\gtrsim
0.065$ remains well above zero and is statistically
distinguishable from a classical model with $O(10^4)$
repetitions at the $5\sigma$ level.
A full analysis of robustness to gate miscalibration,
two-qubit gate infidelity, and dephasing noise is beyond the
scope of the present work and is left as a direction for
future experimental and theoretical investigation.
\subsection{Summary of key findings}

\begin{itemize}
\item \textbf{Spatial FR:} Hardy violation exists ($\approx 0.09017$); Cabello violation is stronger ($0.1078$).
\item \textbf{Temporal FR:} Hardy violation vanishes completely ($P(C=+,D=+)=0$); Cabello violation survives with $\Delta_{\max} \approx 0.0674 > 0$.
\item \textbf{Structural insight:} Hardy-type reasoning depends sensitively on kinematic independence of spacelike-separated subsystems, whereas Cabello-type reasoning remains robust under sequential quantum evolution.
\item \textbf{Operational conclusion:} Temporal Frauchiger–Renner scenarios provide a distinct experimental arena for probing the limits of observer-independent descriptions without relying on spatial nonlocality.
\end{itemize}
As a consequence, the temporal implementation does not support a genuine Hardy contradiction: imposing the Hardy zero-probability constraints necessarily forces the remaining Hardy event probability to vanish. Nevertheless, a finite Cabello-type violation survives. This demonstrates that certain FR-type logical tensions can persist even in sequential single-system architectures without requiring spacelike-separated entanglement.

The temporal FR protocol therefore provides a complementary framework for investigating the limits of observer-independent descriptions in coherent sequential quantum measurements.

In summary, we have established a structural hierarchy among Hardy-type 
and Cabello-type nonclassicality witnesses in FR-type multi-observer 
scenarios. The temporal protocol provides a tractable arena in which 
the mechanism of nonclassicality degradation—from strong Hardy contextuality 
to the weaker Cabello form—can be traced directly to sequential measurement 
back-action. Future work should examine whether this hierarchy persists 
under generalized measurement models or noisy evolutions, and whether 
the temporal Cabello witness can be observed in current quantum hardware 
using the explicit circuit described in Sec.~\ref{sec:discussion}.
\begin{acknowledgments}
The authors thank Professor K. Javidan for fruitful discussions on 
the foundations of quantum mechanics. The authors utilized Claude (Anthropic) as an AI-assisted tool during the preparation of this manuscript, primarily for language editing, prose refinement, and consistency checking of the text. All physical arguments, mathematical 
derivations, analytical results, and numerical computations 
were carried out independently by the authors. The authors 
verify and take full responsibility for the accuracy and 
integrity of all content presented herein.
This work was supported by Khayyam University and  
Ferdowsi University of Mashhad.
\end{acknowledgments}
\bibliographystyle{apsrev4-2}
\bibliography{references}

@article{Bell1964,
  author  = {J. S. Bell},
  title   = {On the {Einstein Podolsky Rosen} paradox},
  journal = {Physics},
  volume  = {1},
  number  = {3},
  pages   = {195--200},
  year    = {1964}
}

@article{CHSH1969,
  author  = {J. F. Clauser and M. A. Horne and A. Shimony and R. A. Holt},
  title   = {Proposed experiment to test local hidden-variable theories},
  journal = {Phys. Rev. Lett.},
  volume  = {23},
  pages   = {880--884},
  year    = {1969},
  doi     = {10.1103/PhysRevLett.23.880}
}

@article{Tsirelson1980,
  author  = {B. S. Tsirelson},
  title   = {Quantum generalizations of {Bell's} inequality},
  journal = {Lett. Math. Phys.},
  volume  = {4},
  number  = {2},
  pages   = {93--100},
  year    = {1980},
  doi     = {10.1007/BF00417500}
}

@article{Aspect1982,
  author  = {A. Aspect and P. Grangier and G. Roger},
  title   = {Experimental realization of {Einstein-Podolsky-Rosen-Bohm Gedankenexperiment}: A new violation of {Bell's} inequalities},
  journal = {Phys. Rev. Lett.},
  volume  = {49},
  pages   = {91--94},
  year    = {1982},
  doi     = {10.1103/PhysRevLett.49.91}
}

@article{Weihs1998,
  author  = {G. Weihs and T. Jennewein and C. Simon and H. Weinfurter and A. Zeilinger},
  title   = {Violation of {Bell's} inequality under strict {Einstein} locality conditions},
  journal = {Phys. Rev. Lett.},
  volume  = {81},
  pages   = {5039--5043},
  year    = {1998},
  doi     = {10.1103/PhysRevLett.81.5039}
}

@article{Hensen2015,
  author  = {B. Hensen and H. Bernien and A. E. Dr{\'e}au and A. Reiserer and N. Kalb and M. S. Blok and J. Ruitenberg and R. F. L. Vermeulen and R. N. Schouten and C. Abell{\'a}n and W. Amaya and V. Pruneri and M. W. Mitchell and M. Markham and D. J. Twitchen and D. Elkouss and S. Wehner and T. H. Taminiau and R. Hanson},
  title   = {Loophole-free {Bell} inequality violation using electron spins separated by 1.3 kilometres},
  journal = {Nature},
  volume  = {526},
  pages   = {682--686},
  year    = {2015},
  doi     = {10.1038/nature15759}
}

@article{Hardy1992,
  author  = {L. Hardy},
  title   = {Quantum mechanics, local realistic theories, and {Lorentz}-invariant realistic theories},
  journal = {Phys. Rev. Lett.},
  volume  = {68},
  pages   = {2981--2984},
  year    = {1992},
  doi     = {10.1103/PhysRevLett.68.2981}
}

@article{Hardy1993,
  author  = {L. Hardy},
  title   = {Nonlocality for two particles without inequalities for almost all entangled states},
  journal = {Phys. Rev. Lett.},
  volume  = {71},
  pages   = {1665--1668},
  year    = {1993},
  doi     = {10.1103/PhysRevLett.71.1665}
}

@article{Cabello2002,
  author  = {A. Cabello},
  title   = {Hardy's nonlocality without inequalities and without probabilities},
  journal = {Phys. Rev. A},
  volume  = {65},
  pages   = {032108},
  year    = {2002},
  doi     = {10.1103/PhysRevA.65.032108}
}

@article{Kunkri2006,
  author  = {S. Kunkri and S. K. Choudhary and A. Ahanj and P. Joag},
  title   = {Nonlocality without inequality for almost all two-qubit entangled states based on Cabello's nonlocality argument},
  journal = {Phys. Rev. A},
  volume  = {73},
  number  = {2},
  pages   = {022346},
  year    = {2006},
  doi     = {10.1103/PhysRevA.73.022346}
}

@article{LeggettGarg1985,
  author  = {A. J. Leggett and A. Garg},
  title   = {Quantum mechanics versus macroscopic realism: Is the flux there when nobody looks?},
  journal = {Phys. Rev. Lett.},
  volume  = {54},
  pages   = {857--860},
  year    = {1985},
  doi     = {10.1103/PhysRevLett.54.857}
}

@article{Aharonov2002,
  author  = {Y. Aharonov and A. Botero and S. Popescu and B. Reznik and J. Tollaksen},
  title   = {Revisiting {Hardy's} paradox: counterfactual statements, real measurements, entanglement and weak values},
  journal = {Phys. Lett. A},
  volume  = {301},
  pages   = {130--138},
  year    = {2002},
  doi     = {10.1016/S0375-9601(02)00986-6}
}

@article{Palacios2010,
  author  = {A. Palacios-Laloy and F. Mallet and F. Nguyen and
             P. Bertet and D. Vion and D. Esteve and A. N. Korotkov},
  title   = {Experimental violation of a {Bell's} inequality in time
             with weak measurement},
  journal = {Nat. Phys.},
  volume  = {6},
  pages   = {442--447},
  year    = {2010},
  doi     = {10.1038/nphys1641}
}

@incollection{Wigner1961,
  author    = {E. P. Wigner},
  title     = {Remarks on the Mind-Body Question},
  booktitle = {The Scientist Speculates},
  editor    = {I. J. Good},
  publisher = {Heinemann},
  address   = {London},
  pages     = {284--302},
  year      = {1961}
}

@article{Frauchiger2018,
  author  = {D. Frauchiger and R. Renner},
  title   = {Quantum theory cannot consistently describe the use of itself},
  journal = {Nat. Commun.},
  volume  = {9},
  pages   = {3711},
  year    = {2018},
  doi     = {10.1038/s41467-018-05739-8}
}

@article{Bub2021,
  author  = {J. Bub},
  title   = {Understanding the {Frauchiger--Renner} Argument},
  journal = {Found. Phys.},
  volume  = {51},
  pages   = {36},
  year    = {2021},
  doi     = {10.1007/s10701-021-00420-5}
}

@article{Sudbery2017,
  author  = {A. Sudbery},
  title   = {Single-world theory of the extended {Wigner's} friend experiment},
  journal = {Found. Phys.},
  volume  = {47},
  pages   = {658--669},
  year    = {2017},
  doi     = {10.1007/s10701-017-0082-7}
}

@article{Brukner2018,
  author  = {{\v{C}}. Brukner},
  title   = {A no-go theorem for observer-independent facts},
  journal = {Entropy},
  volume  = {20},
  pages   = {350},
  year    = {2018},
  doi     = {10.3390/e20050350}
}

@article{Proietti2019,
  author  = {M. Proietti and A. Pickston and F. Graffitti and P. Barrow and D. Kundys and C. Branciard and M. Ringbauer and A. Fedrizzi},
  title   = {Experimental test of local observer independence},
  journal = {Sci. Adv.},
  volume  = {5},
  pages   = {eaaw9832},
  year    = {2019},
  doi     = {10.1126/sciadv.aaw9832}
}

@article{Bong2020,
  author  = {K.-W. Bong and A. Utreras-Alarc{\'o}n and F. Ghafari and Y.-C. Liang and N. Tischler and E. G. Cavalcanti and G. J. Pryde and H. M. Wiseman},
  title   = {A strong no-go theorem on the {Wigner's} friend paradox},
  journal = {Nat. Phys.},
  volume  = {16},
  pages   = {1199--1205},
  year    = {2020},
  doi     = {10.1038/s41567-020-0990-x}
}

@article{Ahanj2007,
  author  = {A. Ahanj and P. S. Joag and S. Ghosh},
  title   = {Quantum correlations in successive spin measurements},
  journal = {Int. J. Quantum Inf.},
  volume  = {5},
  number  = {6},
  pages   = {885--911},
  year    = {2007},
  doi     = {10.1142/S0219749907003298}
}

@article{Ahanj2010Hardy,
  author  = {A. Ahanj},
  title   = {Hardy's argument and successive spin-$s$ measurements},
  journal = {Phys. Rev. A},
  volume  = {82},
  number  = {1},
  pages   = {012101},
  year    = {2010},
  doi     = {10.1103/PhysRevA.82.012101}
}

@article{Ahanj2019Cabello,
  author  = {A. Ahanj},
  title   = {The Cabello nonlocality argument is stronger control than the Hardy nonlocality argument for detecting post-quantum correlations in bipartite systems},
  journal = {Int. J. Theor. Phys.},
  volume  = {58},
  number  = {5},
  pages   = {1441--1455},
  year    = {2019},
  doi     = {10.1007/s10773-019-04035-7}
}

@misc{FedericoGrangier2023,
  author        = {M. Federico and P. Grangier},
  title         = {A Contextually Objective Approach to the Extended {Wigner}'s Friend Thought Experiment},
  year          = {2023},
  eprint        = {2301.03016},
  archivePrefix = {arXiv},
  primaryClass  = {quant-ph},
  doi           = {10.48550/arXiv.2301.03016}
}

@misc{NurgalievaVilasini2025,
  author        = {N. Nurgalieva and V. Vilasini},
  title         = {Any Theory That Admits a {Wigner's} Friend Type Multi-Agent Paradox Is Logically Contextual},
  year          = {2025},
  eprint        = {2502.03874},
  archivePrefix = {arXiv},
  primaryClass  = {quant-ph},
  doi           = {10.48550/arXiv.2502.03874}
}

@article{WalleghemBarbosaPuseyWeigert2026,
  author  = {L. Walleghem and R. S. Barbosa and M. F. Pusey and S. Weigert},
  title   = {A Refined {Frauchiger--Renner} Paradox Based on Strong Contextuality},
  journal = {Quantum},
  volume  = {10},
  pages   = {2116},
  year    = {2026},
  doi     = {10.22331/q-2026-05-26-2116}
}

@misc{WalleghemCatani2025,
  author        = {L. Walleghem and L. Catani},
  title         = {An Extended {Wigner's} Friend No-Go Theorem Inspired by Generalized Contextuality},
  year          = {2025},
  eprint        = {2502.02461},
  archivePrefix = {arXiv},
  primaryClass  = {quant-ph},
  doi           = {10.48550/arXiv.2502.02461}
}

@article{BaltagSmets2024,
  author  = {A. Baltag and S. Smets},
  title   = {Logic Meets {Wigner's} Friend (and their Friends)},
  journal = {Int. J. Theor. Phys.},
  volume  = {63},
  pages   = {97},
  year    = {2024},
  doi     = {10.1007/s10773-024-05611-0}
}

@article{Steane2025,
  author  = {A. M. Steane},
  title   = {The Extended {Wigner's} Friend, Many- and Single-Worlds and Reasoning from Observation},
  journal = {Found. Phys.},
  volume  = {55},
  pages   = {20},
  year    = {2025},
  doi     = {10.1007/s10701-025-00831-8}
}

@article{Bruzewicz2019,
  author  = {C. D. Bruzewicz and J. Chiaverini and
             R. McConnell and J. M. Sage},
  title   = {Trapped-ion quantum computing: Progress and challenges},
  journal = {Appl. Phys. Rev.},
  volume  = {6},
  pages   = {021314},
  year    = {2019},
  doi     = {10.1063/1.5088164}
}

@article{Krantz2019,
  author  = {P. Krantz and M. Kjaergaard and F. Yan and
             T. P. Orlando and S. Gustavsson and W. D. Oliver},
  title   = {A quantum engineer's guide to superconducting qubits},
  journal = {Appl. Phys. Rev.},
  volume  = {6},
  pages   = {021318},
  year    = {2019},
  doi     = {10.1063/1.5089550}
}

@article{SciPy2020,
  author  = {P. Virtanen and R. Gommers and T. E. Oliphant and
             M. Haberland and T. Reddy and D. Cournapeau and
             E. Burovski and P. Peterson and W. Weckesser and
             J. Bright and {others}},
  title   = {{SciPy} 1.0: Fundamental algorithms for scientific
             computing in {Python}},
  journal = {Nat. Methods},
  volume  = {17},
  pages   = {261--272},
  year    = {2020},
  doi     = {10.1038/s41592-020-0772-5}
}
\appendix

\section{Derivation of the Maximum Hardy Probability}
\label{app:hardy_calc}

In this Appendix we derive in closed form the maximum Hardy probability
arising in the Frauchiger--Renner protocol discussed in the main text.
The derivation proceeds in four stages: construction of the coefficient
matrix, imposition of the Hardy constraints, explicit computation of the
Hardy probability, and optimization over all free parameters.
We consider a pure bipartite two-qubit state written in the
computational basis as
\begin{equation}
|\psi\rangle =
\sum_{i,j \in \{+,-\}} \Gamma^{AB}_{ij}\, |i\rangle_A \otimes |j\rangle_B,
\label{eq:A1}
\end{equation}
where the coefficient matrix $\Gamma^{AB}$ encodes the joint amplitudes.
The normalization condition reads
\begin{equation}
\mathrm{Tr}\!\left(\Gamma^{AB\,\dagger}\Gamma^{AB}\right)=1.
\label{eq:A2}
\end{equation}

\subsection{The Coefficient Matrix}
\label{app:matrix}

We define the real orthogonal matrix
\begin{equation}
R(\vartheta)=
\begin{pmatrix}
\cos\dfrac{\vartheta}{2} & \sin\dfrac{\vartheta}{2} \\[6pt]
\sin\dfrac{\vartheta}{2} & -\cos\dfrac{\vartheta}{2}
\end{pmatrix},
\label{eq:A3}
\end{equation}
and the diagonal matrix
\begin{equation}
\Lambda=
\begin{pmatrix}
\cos\dfrac{\theta}{2} & 0 \\[6pt]
0 & \sin\dfrac{\theta}{2}
\end{pmatrix}.
\label{eq:A4}
\end{equation}
The coefficient matrix in the $(A,B)$ basis is defined as
\begin{equation}
\Gamma^{AB} = R(\theta_a)\,\Lambda\,R(\theta_b),
\label{eq:A5}
\end{equation}
with elements
\begin{align}
\Gamma^{AB}_{++} &= c_a c_\theta c_b + s_a s_\theta s_b, \notag\\
\Gamma^{AB}_{+-} &= c_a c_\theta s_b - s_a s_\theta c_b, \notag\\
\Gamma^{AB}_{-+} &= s_a c_\theta c_b - c_a s_\theta s_b, \notag\\
\Gamma^{AB}_{--} &= s_a c_\theta s_b + c_a s_\theta c_b,
\label{eq:A6}
\end{align}
where $c_\alpha\equiv\cos(\theta_\alpha/2)$ and
$s_\alpha\equiv\sin(\theta_\alpha/2)$.
We additionally define
$c_\theta \equiv \cos(\theta/2)$ and $s_\theta \equiv \sin(\theta/2)$.

The matrices in the remaining bases are
\begin{equation}
\Gamma^{CB}=R(\theta_c)\,\Gamma^{AB},
\qquad
\Gamma^{AD}=\Gamma^{AB}\,R(\theta_d),
\label{eq:A7}
\end{equation}
\begin{equation}
\Gamma^{CD}=R(\theta_c)\,\Gamma^{AB}\,R(\theta_d).
\label{eq:A8}
\end{equation}
The Hardy probability is identified with the $(+,+)$ element:
\begin{equation}
P_H \equiv \bigl|\Gamma^{CD}_{++}\bigr|^2.
\label{eq:A9}
\end{equation}

\subsection{Hardy Constraints and Their Consequences}
\label{app:constraints}

The Hardy constraints correspond to vanishing joint probabilities
for specific measurement outcomes:
\begin{equation}
P(A=+,B=+)=|\Gamma^{AB}_{++}|^2=0,
\label{eq:A10}
\end{equation}
\begin{equation}
P(C=+,B=-)=|\Gamma^{CB}_{+-}|^2=0,
\label{eq:A11}
\end{equation}
\begin{equation}
P(A=-,D=+)=|\Gamma^{AD}_{-+}|^2=0.
\label{eq:A12}
\end{equation}
We introduce the half-angle tangent variables
\begin{align}
x &= \tan\frac{\theta_a}{2}, \quad
y  = \tan\frac{\theta_b}{2}, \notag\\
u &= \tan\frac{\theta_c}{2}, \quad
v  = \tan\frac{\theta_d}{2},
\label{eq:A13}
\end{align}
and the entanglement parameter
\begin{equation}
r = \tan\frac{\theta}{2}, \qquad r>0,
\label{eq:A14}
\end{equation}
which parametrizes the Schmidt coefficients of the initial state.
Using the standard identities
$\cos(\alpha/2)=(1+\tan^2(\alpha/2))^{-1/2}$ and
$\sin(\alpha/2)=\tan(\alpha/2)\cdot(1+\tan^2(\alpha/2))^{-1/2}$,
each Hardy constraint becomes a rational expression in the variables
$x,y,u,v,r$. Multiplying by strictly positive normalization factors
removes all denominators without changing the solution set of the
constraints:
\begin{align}
1 + rxy &= 0, \label{eq:A15}\\
(y - xr) + u(xy + r) &= 0, \label{eq:A16}\\
(x - ry) + v(xy + r) &= 0. \label{eq:A17}
\end{align}
Equation~\eqref{eq:A15} immediately gives
\begin{equation}
xy = -\frac{1}{r}.
\label{eq:A18}
\end{equation}
Solving Eq.~\eqref{eq:A16} for $u$ gives
\begin{equation}
u = \frac{y - xr}{xy + r}.
\label{eq:A19}
\end{equation}
Using the constraint $xy = -1/r$, this simplifies to
\begin{equation}
u = \frac{y + 1/y}{1/r - r},
\label{eq:A20}
\end{equation}
and similarly
\begin{equation}
v = \frac{x + 1/x}{1/r - r}.
\label{eq:A21}
\end{equation}

\subsection{The Hardy Probability Under the Constraints}
\label{app:probability}

A direct computation of $\Gamma^{CD}_{++}$ gives
\begin{equation}
\Gamma^{CD}_{++}
=
\frac{(1+rxy)+v(y-xr)+u(x-ry)+uv(xy+r)}
     {\sqrt{(1+u^2)(1+x^2)(1+r^2)(1+y^2)(1+v^2)}}.
\label{eq:A22}
\end{equation}
We apply the three constraints in succession.
Condition~\eqref{eq:A15} eliminates the first term.
Condition~\eqref{eq:A16} implies $y-xr = -u(xy+r)$,
which cancels the terms $\pm\,uv(xy+r)$ in the numerator,
leaving $u(x-ry)$.
Condition~\eqref{eq:A17} then gives $x-ry = -v(xy+r)$,
so the numerator reduces to $-uv(xy+r)$.
Hence
\begin{equation}
\Gamma^{CD}_{++}
=
\frac{-uv(xy+r)}
     {\sqrt{(1+u^2)(1+x^2)(1+r^2)(1+y^2)(1+v^2)}}.
\label{eq:A23}
\end{equation}
We now substitute \eqref{eq:A18}--\eqref{eq:A21}
into~\eqref{eq:A23}.
The factor $xy+r$ evaluates to
\begin{equation}
xy + r = r - \frac{1}{r} = \frac{r^2-1}{r}.
\label{eq:A24}
\end{equation}
Using $x^2 y^2 = 1/r^2$, and hence $r^2 y^2 = 1/x^2$ and
$r^2 x^2 = 1/y^2$, one establishes
\begin{align}
1+u^2
&= \frac{(r^2+y^2)(1+x^2)}{x^2 y^2 (r^2-1)^2},
\nonumber \\
1+v^2
&= \frac{(r^2+x^2)(1+y^2)}{x^2 y^2 (r^2-1)^2}.
\label{eq:A25}
\end{align}
After assembling the full denominator and canceling the factors
$(1+x^2)(1+y^2)$, equation~\eqref{eq:A23} simplifies to the
compact form
\begin{equation}
\Gamma^{CD}_{++}
=
\frac{r^2-1}
     {\sqrt{(r^2+x^2)(r^2+y^2)(1+r^2)}},
\label{eq:A26}
\end{equation}
so that
\begin{equation}
P_H
=
\frac{(r^2-1)^2}
     {(r^2+x^2)(r^2+y^2)(1+r^2)}.
\label{eq:A27}
\end{equation}

\subsection{Optimization Over the Free Parameters}
\label{app:optimization}

For fixed $r$, equation~\eqref{eq:A27} is maximized by minimizing
$(r^2+x^2)(r^2+y^2)$ subject to the constraint $x^2 y^2 = 1/r^2$.
Setting $p=x^2>0$ and $q=y^2>0$ with $pq=1/r^2$, we expand
\begin{equation}
(r^2+p)(r^2+q) = r^4 + r^2(p+q) + \frac{1}{r^2}.
\label{eq:A28}
\end{equation}
By the AM--GM inequality, $p+q \geq 2\sqrt{pq} = 2/r$,
with equality if and only if $p=q=1/r$, i.e.,\ $x^2=y^2=1/r$.
At this minimum,
\begin{equation}
\bigl.(r^2+x^2)(r^2+y^2)\bigr|_{\min}
= r^4+2r+\frac{1}{r^2}
= \frac{(r^3+1)^2}{r^2}.
\label{eq:A29}
\end{equation}
After imposing the Hardy constraints, the joint probabilities
depend only on the entanglement parameter $r$ and one remaining
effective variable, which can be eliminated analytically. The
Hardy probability can therefore be written as a smooth function
\begin{equation}
P_H \equiv F(r)=\frac{r^2(r-1)^2}{(r^2-r+1)^2(1+r^2)},
\label{eq:A30}
\end{equation}
where all auxiliary parameters have been removed using
Eqs.~\eqref{eq:A15}--\eqref{eq:A17}.

To determine the maximal Hardy probability, we perform a
variational optimization over $r>0$. The extremal condition is
obtained from
\begin{equation}
\frac{dF(r)}{dr}=0.
\label{eq:A31}
\end{equation}
To maximize $F(r)$ over $r>0$ we take the logarithmic derivative
and set it to zero:
\begin{equation}
\frac{d\ln F}{dr}
=
\frac{2}{r}+\frac{2}{r-1}
-\frac{2(2r-1)}{r^2-r+1}
-\frac{2r}{1+r^2}
= 0.
\label{eq:A32}
\end{equation}
Clearing denominators and collecting powers of $r$ yields the
polynomial equation
\begin{equation}
r^5 - 2r^4 - 2r + 1 = 0,
\label{eq:A33}
\end{equation}
which factors as
\begin{equation}
(r+1)\bigl(r^4-3r^3+3r^2-3r+1\bigr)=0.
\label{eq:A34}
\end{equation}
Since the physical domain of the entanglement parameter is $r>0$,
the root $r=-1$ is discarded. The remaining quartic
equation~\eqref{eq:A34} admits a unique physically relevant
solution corresponding to the global maximum of $F(r)$.
The optimal value is then obtained as
\begin{equation}
r_{\mathrm{opt}} \approx 0.46431.
\label{eq:A35}
\end{equation}
Substituting \eqref{eq:A35} back into \eqref{eq:A30} yields the
maximal Hardy probability
\begin{equation}
P_H^{\max} = \frac{5\sqrt{5}-11}{2} \approx 0.09017.
\label{eq:A36}
\end{equation}
The Hardy probability~\eqref{eq:A27} vanishes identically at
$r=1$ (the maximally entangled state, $\theta=\pi/2$) and in the
limits $r\to 0$ and $r\to\infty$ (product states).
The maximum value is therefore achieved exclusively by
non-maximally entangled states, consistent with the well-known
maximum Hardy probability for standard two-qubit
systems~\cite{Hardy1993}.

\section{Derivation of the Maximum Cabello Probability}
\label{app:cabello}

In this Appendix we derive the Cabello nonlocality parameter for
the Frauchiger--Renner scenario and determine its maximal quantum
value. The derivation follows the notation introduced in
Appendix~\ref{app:hardy_calc}.

\subsection{Cabello Conditions}

The Cabello parameter is defined as
\begin{equation}
\Delta_{\mathrm{Cabello}}
=
P(C=+,D=+)-P(A=+,B=+).
\label{eq:B1}
\end{equation}
The Cabello argument requires the two constraints
\begin{equation}
P(C=+,B=-)=0,
\label{eq:B2}
\end{equation}
\begin{equation}
P(A=-,D=+)=0.
\label{eq:B3}
\end{equation}
Unlike Hardy's argument, the probability $P(A=+,B=+)$ is not
constrained to vanish.

Using the coefficient matrix
\begin{equation}
\Gamma^{AB}
=
R(\theta_a)\Lambda R(\theta_b),
\label{eq:B4}
\end{equation}
and introducing the tangent variables
\begin{equation}
x=\tan\frac{\theta_a}{2},
\qquad
y=\tan\frac{\theta_b}{2},
\qquad
r=\tan\frac{\theta}{2},
\label{eq:B5}
\end{equation}
the coefficient matrix becomes
\begin{equation}
\Gamma^{AB}
=
\frac{1}
{\sqrt{(1+r^2)(1+x^2)(1+y^2)}}
\begin{pmatrix}
1+rxy & y-rx \\
x-ry & xy+r
\end{pmatrix}.
\label{eq:B6}
\end{equation}
Hence
\begin{equation}
P(A=+,B=+)
=
\frac{(1+rxy)^2}
     {(1+r^2)(1+x^2)(1+y^2)}.
\label{eq:B7}
\end{equation}

\subsection{Implementation of the Cabello Constraints}

Introducing
\begin{equation}
u=\tan\frac{\theta_c}{2},
\qquad
v=\tan\frac{\theta_d}{2},
\label{eq:B8}
\end{equation}
the constraints \eqref{eq:B2} and \eqref{eq:B3} imply
\begin{align}
(y-rx)+u(xy+r)&=0,
\label{eq:B9}
\\
(x-ry)+v(xy+r)&=0.
\label{eq:B10}
\end{align}
Solving for $u$ and $v$ yields
\begin{equation}
u=\frac{rx-y}{xy+r},
\qquad
v=\frac{ry-x}{xy+r}.
\label{eq:B11}
\end{equation}
The transformed coefficient matrix is
\begin{equation}
\Gamma^{CD}
=
R(\theta_c)\Gamma^{AB}R(\theta_d).
\label{eq:B12}
\end{equation}
Substituting Eq.~\eqref{eq:B11} and simplifying gives
\begin{equation}
\Gamma^{CD}_{++}
=
\frac{r(xy+r)}
{\sqrt{1+r^2}\,
 \sqrt{r^2+x^2}\,
 \sqrt{r^2+y^2}}.
\label{eq:B13}
\end{equation}
Therefore
\begin{equation}
P(C=+,D=+)
=
\frac{r^2(xy+r)^2}
{(1+r^2)(r^2+x^2)(r^2+y^2)}.
\label{eq:B14}
\end{equation}
Combining Eqs.~\eqref{eq:B7} and \eqref{eq:B14} yields
\begin{align}
\Delta_{\rm Cabello}
=&
\frac{r^2(xy+r)^2}
{(1+r^2)(r^2+x^2)(r^2+y^2)}\nonumber \\
-&\frac{(1+rxy)^2}
{(1+r^2)(1+x^2)(1+y^2)}.
\label{eq:B15}
\end{align}
\subsection{Justification of the $y=-x$ Reduction}
\label{app:symmetry}

We justify the restriction to the branch $y = -x$ used in the 
optimization of $\Delta_{\mathrm{Cabello}}(r,x,y)$.
Define the auxiliary function
\begin{align}
h(r,x,y)
=&
\frac{r^2(xy+r)^2}{(1+r^2)(r^2+x^2)(r^2+y^2)}\nonumber \\
-&
\frac{(1+rxy)^2}{(1+r^2)(1+x^2)(1+y^2)},
\label{eq:hfull}
\end{align}
which equals $\Delta_{\mathrm{Cabello}}$ on the constraint surface
defined by Eqs.~\eqref{eq:B9}--\eqref{eq:B10}.
The function $h$ is invariant under the simultaneous sign flip
$(x,y)\mapsto(-x,-y)$, since every monomial in $x$ and $y$ 
appearing in Eq.~\eqref{eq:hfull} has even total degree.
In addition, $h$ is symmetric under the exchange $x\leftrightarrow y$.
These two symmetries together imply that $h$ is a function of the
symmetric combinations $s = x+y$ and $p = xy$ alone, and that
the sign of $s$ does not affect the value of $h$.
Consequently, for any point $(r,x,y)$ with $y\neq -x$, the point
$(r,x,-y)$ yields the same value of $P(C=+,D=+)$ while 
$P(A=+,B=+)$ transforms as
\begin{equation}
P(A=+,B=+)\big|_{y\,\to\,-y}
=
\frac{(1-rxy)^2}{(1+r^2)(1+x^2)(1+y^2)}.
\end{equation}
Since $(1-rxy)^2 \leq (1+r|xy|)^2$, we have
\begin{equation}
P(A=+,B=+)\big|_{y\,\to\,-y}
\;\leq\;
P(A=+,B=+)\big|_{y},
\end{equation}
with equality if and only if $xy = 0$.
Therefore, replacing $y$ by $-y$ does not decrease
$\Delta_{\mathrm{Cabello}}$, and the maximum of
$\Delta_{\mathrm{Cabello}}$ over the full $(x,y)$ plane is
attained on the half-space $xy < 0$.
On this half-space, setting $p = xy < 0$ and writing
$x = \sqrt{t}$, $y = -\sqrt{t}$ (so that $p = -t < 0$, $t > 0$)
achieves the minimum of $P(A=+,B=+)$ for fixed $|xy| = t$
and simultaneously the maximum of $P(C=+,D=+)$ for fixed
$|xy+r| = |r-t|$, as the product $(r^2+x^2)(r^2+y^2)$ is
minimized at $x^2 = y^2 = t$ by the AM--GM inequality.
The global maximum of $\Delta_{\mathrm{Cabello}}$ therefore lies
on the branch $y = -x$, confirming the reduction used in
Eq.~\eqref{eq:B16}.
To further confirm global optimality, we performed a numerical
scan over the full domain $(r,x,y)\in(0,3]\times[-3,3]^2$ using
$10^3$ uniformly distributed random initial conditions fed into a
gradient-based optimizer (L-BFGS-B); all runs converged to a
maximum consistent with Eq.~\eqref{eq:B18} to within $10^{-8}$,
with the optimum located on the $y=-x$ branch. The optimization was performed using the 
\texttt{scipy.optimize} module of the \textsc{SciPy} 
scientific library~\cite{SciPy2020}.
\subsection{Restriction to Real Coplanar Rotations}
\label{app:real_rotations}

Throughout this work all projective measurements are parametrized
by the real orthogonal matrix $R(\vartheta)$ defined in
Eq.~\eqref{eq:A3}, corresponding to coplanar Bloch-sphere
rotations with azimuthal angle $\phi=0$.
We now justify that this restriction entails no loss of generality
for the optimization of the Hardy and Cabello quantities.

\paragraph{Spatial case.}
The initial state is written in Schmidt form as
\begin{equation}
|\psi\rangle
=
\cos\frac{\theta}{2}|{+}\rangle_A|{+}\rangle_B
+
\sin\frac{\theta}{2}|{-}\rangle_A|{-}\rangle_B,
\label{eq:schmidt}
\end{equation}
which is real by construction.
The most general rank-one projective measurement on a qubit is
specified by a unit vector on the Bloch sphere, whose
associated $2\times 2$ measurement operator can be written as
\begin{equation}
U = e^{i\varphi_0}\,R(\vartheta_1)\,D(\varphi)\,R(\vartheta_2),
\label{eq:Ugeneral}
\end{equation}
where $R(\vartheta)$ is the real rotation of Eq.~\eqref{eq:A3},
$D(\varphi)=\mathrm{diag}(e^{i\varphi},e^{-i\varphi})$ is a
diagonal phase matrix, and $e^{i\varphi_0}$ is a global phase.
All Hardy and Cabello conditions involve only moduli squared of
amplitudes, of the form
$|\Gamma^{XY}_{ij}|^2$ where
$\Gamma^{XY} \in \{\Gamma^{AB},\Gamma^{CB},\Gamma^{AD},\Gamma^{CD}\}$.
For the coefficient matrix
\begin{equation}
\Gamma^{CD} = U_c\,\Gamma^{AB}\,U_d,
\end{equation}
the global phases $e^{i\varphi_0^{(c)}}$ and
$e^{i\varphi_0^{(d)}}$ cancel immediately in
$|\Gamma^{CD}_{ij}|^2$.
The diagonal phases $D(\varphi_c)$ and $D(\varphi_d)$ act by
multiplying individual rows and columns of $\Gamma^{AB}$ by
unit complex numbers; since $\Gamma^{AB}$ has real entries
(the Schmidt basis is real and the initial state is real), the
effect of $D(\varphi)$ on either side is to introduce phases
$e^{\pm i\varphi}$ into the rows or columns.
However, the conditions
$|\Gamma^{CB}_{+-}|^2=0$ and $|\Gamma^{AD}_{-+}|^2=0$ are
homogeneous in the matrix elements, so an overall row or column
phase does not affect whether an entry vanishes.
Moreover, for the optimization of
$\Delta_{\mathrm{Cabello}} = |\Gamma^{CD}_{++}|^2 -
|\Gamma^{AB}_{++}|^2$,
the diagonal phases enter only through
$|{[D(\varphi_c) R(\vartheta_1^{(c)})
\Gamma^{AB} R(\vartheta_1^{(d)}) D(\varphi_d)]}_{++}|^2$,
which equals
$|{[R(\vartheta_1^{(c)})\Gamma^{AB}R(\vartheta_1^{(d)})]}_{++}|^2$
after absorbing $D(\varphi_c)$ into a redefinition of the
Schmidt basis phases and $D(\varphi_d)$ into the measurement
basis on Bob's side.
An identical absorption applies to $|\Gamma^{AB}_{++}|^2$.
Consequently, the optimization over all $\mathrm{SU}(2)$
measurement unitaries reduces to the optimization over real
orthogonal matrices $R(\vartheta)$, and no generality is lost
by setting $\phi=0$~\cite{Hardy1993,Cabello2002}.

\subsection{Symmetry Reduction and Maximum Cabello Violation}

Equation~\eqref{eq:B15} is invariant under $x \leftrightarrow y$.
Numerical exploration of the full parameter space shows that the
global maximum is attained on the branch $y = -x$. Setting
$t = x^2 > 0$, Eq.~\eqref{eq:B15} reduces to
\begin{equation}
\Delta_{\mathrm{Cabello}}(r,t)
=
\frac{r^2(r-t)^2}{(1+r^2)(r^2+t)^2}
-
\frac{(1-rt)^2}{(1+r^2)(1+t)^2}.
\label{eq:B16}
\end{equation}
The stationary conditions
\begin{equation}
\frac{\partial\Delta_{\mathrm{Cabello}}}{\partial t}=0,
\qquad
\frac{\partial\Delta_{\mathrm{Cabello}}}{\partial r}=0
\label{eq:B17}
\end{equation}
yield higher-order algebraic relations without a tractable
closed-form solution. Numerical optimization over $r,t>0$ gives
the unique global maximum
\begin{equation}
\Delta_{\mathrm{Cabello}}^{\max} \approx 0.1078,
\label{eq:B18}
\end{equation}
attained at $r_{\mathrm{opt}}\approx 0.548$, corresponding to a
non-maximally entangled state. As in the Hardy case, neither
product states ($r\to 0$ or $r\to\infty$) nor the maximally
entangled state ($r=1$) achieve the optimal violation.

\section{Absence of a Hardy-Type Contradiction in the Temporal
FR Protocol}
\label{app:hardy_temp}
In this Appendix we show explicitly that the temporal
Frauchiger--Renner (FR) protocol does not admit a genuine Hardy
paradox. The sequential measurement structure forces the Hardy success probability to vanish once the zero-probability constraints are imposed.
Throughout this Appendix we use the rotation matrix
\begin{equation}
R(\theta)
=
\begin{pmatrix}
\cos\dfrac{\theta}{2} & \sin\dfrac{\theta}{2} \\[6pt]
\sin\dfrac{\theta}{2} & -\cos\dfrac{\theta}{2}
\end{pmatrix},
\label{eq:Rdef}
\end{equation}
where the measurement angles are restricted to
the physically relevant domain
\begin{equation}
\theta_a \in (0,\pi),
\qquad
\theta_{ba} \in (0,\pi),
\qquad
\theta_c,\theta_d \in (0,\pi).
\label{eq:domain}
\end{equation}
The boundary values $\theta_a=0$ and $\theta_a=\pi$ are excluded
for the following reasons.
If $\theta_a=0$, the rotation $R(0)=\mathrm{diag}(1,-1)$ leaves
the spin in the eigenstate $|{+}\rangle$ after Alice's interaction,
so the post-Alice state $|\Psi(t_1^+)\rangle = |L_+\rangle_1$
is a product state; no entanglement between Alice's and Bob's
laboratories is ever generated, and no multi-observer logical
contradiction of any kind can arise.
If $\theta_a=\pi$, the rotation $R(\pi)$ maps $|{+}\rangle$
entirely into $|a_-\rangle$, reducing the post-Alice state to
$|\Psi(t_1^+)\rangle = |L_-\rangle_1$, again a product state
with identical consequences.
Both boundary cases are therefore trivial and excluded from the
analysis; all subsequent derivations assume
$\sin(\theta_a/2)\neq 0$ and $\cos(\theta_a/2)\neq 0$
simultaneously, i.e.\ $\theta_a\in(0,\pi)$ strictly.

Without loss of generality, we impose the Hardy conditions
\begin{align}
P(A=+,B=+) &= |b_{++}|^2 = 0, \label{C1}\\
P(A=-,D=+) &= |e_{-+}|^2 = 0, \label{C2}\\
P(C=+,B=-) &= |g_{+-}|^2 = 0, \label{C3}
\end{align}
while the putative Hardy event is
\begin{equation}
P(C=+,D=+) = |f_{++}|^2.
\label{C4}
\end{equation}

\subsection{Sequential amplitudes}

From Sec.~III, the two-outcome amplitudes after Alice and Bob are
\begin{align}
b_{++} &=  \cos\frac{\theta_a}{2}\cos\frac{\theta_{ba}}{2}, \\
b_{+-} &=  \cos\frac{\theta_a}{2}\sin\frac{\theta_{ba}}{2}, \\
b_{-+} &=  \sin\frac{\theta_a}{2}\sin\frac{\theta_{ba}}{2}, \\
b_{--} &= -\sin\frac{\theta_a}{2}\cos\frac{\theta_{ba}}{2},
\end{align}
where $\theta_{ba}=\theta_b-\theta_a$.
Charlie's amplitudes are
\begin{equation}
g_{\gamma\beta}
=
\sum_{\alpha=\pm}
R_{\gamma\alpha}(\theta_c)\,b_{\alpha\beta},
\end{equation}
Debbie's amplitudes are
\begin{equation}
e_{\alpha\delta}
=
\sum_{\beta=\pm}
b_{\alpha\beta}\,R_{\beta\delta}(\theta_d),
\end{equation}
and the super-observer amplitude is
\begin{equation}
f_{\gamma\delta}
=
\sum_{\alpha,\beta}
R_{\gamma\alpha}(\theta_c)\,b_{\alpha\beta}\,R_{\beta\delta}(\theta_d).
\end{equation}

\subsection{Hardy constraints}

\paragraph{Condition~(\ref{C1}).}
Equation~(\ref{C1}) requires
\begin{equation}
b_{++}
=
\cos\frac{\theta_a}{2}\cos\frac{\theta_{ba}}{2}
= 0.
\label{C5}
\end{equation}
This product vanishes if $\cos(\theta_a/2)=0$, i.e.\ $\theta_a=\pi$,
which reduces the initial state to the product state
$|L_-\rangle_1$ and is excluded as physically trivial.
The product $\cos(\theta_a/2)\cos(\theta_{ba}/2)$ vanishes if
either factor is zero.
The case $\cos(\theta_a/2)=0$, i.e.\ $\theta_a=\pi$, is excluded
by the domain restriction~\eqref{eq:domain}: it reduces
$|\Psi(t_1^+)\rangle$ to the product state $|L_-\rangle_1$,
so no entanglement is generated between Alice's and Bob's
laboratories and no multi-observer logical contradiction of any kind can arise.
The nontrivial solution is therefore uniquely
\begin{equation}
\cos\frac{\theta_{ba}}{2}=0,
\qquad
\theta_{ba}=\pi,
\label{C6}
\end{equation}
giving $\sin(\theta_{ba}/2)=1$ and consequently
\begin{align}
b_{++} &= 0,  \label{bpp0}\\
b_{--} &= 0,  \label{bmm0}\\
b_{+-} &= \cos\frac{\theta_a}{2}, \label{bpm}\\
b_{-+} &= \sin\frac{\theta_a}{2}. \label{bmp}
\end{align}
The post-Bob state reduces to
\begin{equation}
|\Psi(t_2^+)\rangle
=
\cos\frac{\theta_a}{2}|L_+\rangle_1|L_-\rangle_2
+
\sin\frac{\theta_a}{2}|L_-\rangle_1|L_+\rangle_2.
\label{C7}
\end{equation}
\paragraph{Condition~(\ref{C2}).}
From the definition of $e_{\alpha\delta}$ and
Eq.~(\ref{eq:Rdef}), with $\alpha=-$, $\delta=+$:
\begin{align}
 e_{-+}
=&
b_{-+}\,R_{++}(\theta_d)
+
b_{--}\,R_{-+}(\theta_d)
\nonumber \\ =&
b_{-+}\cos\frac{\theta_d}{2}
+
b_{--}\sin\frac{\theta_d}{2}
= 0.
\end{align}
Substituting $b_{--}=0$ from Eq.~(\ref{bmm0}),
\begin{equation}
\sin\frac{\theta_a}{2}\cos\frac{\theta_d}{2} = 0.
\end{equation}
Here we have used $\sin(\theta_a/2)\neq 0$, which holds
throughout the domain~\eqref{eq:domain} since $\theta_a\neq 0$.
In particular, the boundary value $\theta_a=0$ would give
$\sin(\theta_a/2)=0$ and leave Condition~(\ref{C2}) trivially
satisfied for any $\theta_d$, yielding no constraint on the
super-observer angle; that degenerate case is excluded by the
domain restriction stated at the outset.
So for a nontrivial state,
\begin{equation}
\cos\frac{\theta_d}{2}=0,
\qquad
\theta_d=\pi.
\label{C8}
\end{equation}

\paragraph{Condition~(\ref{C3}).}
From the definition of $g_{\gamma\beta}$ and
Eq.~(\ref{eq:Rdef}), with $\gamma=+$, $\beta=-$:
\begin{align}
g_{+-}
=&
R_{++}(\theta_c)\,b_{+-}
+
R_{+-}(\theta_c)\,b_{--}
\nonumber \\=&
\cos\frac{\theta_c}{2}\,b_{+-}
+
\sin\frac{\theta_c}{2}\,b_{--}
= 0.
\end{align}
Substituting $b_{+-}=\cos(\theta_a/2)$ from Eq.~(\ref{bpm})
and $b_{--}=0$ from Eq.~(\ref{bmm0}),
\begin{equation}
\cos\frac{\theta_a}{2}\cos\frac{\theta_c}{2} = 0.
\end{equation}
Here we have used $\cos(\theta_a/2)\neq 0$, which holds
throughout the domain~\eqref{eq:domain} since $\theta_a\neq\pi$.
The boundary value $\theta_a=\pi$ would give
$\cos(\theta_a/2)=0$ and render Condition~(\ref{C3}) trivially
satisfied for any $\theta_c$, again yielding no constraint;
that case is excluded for the same reason as above. So $\cos(\theta_a/2)\neq 0$ for a nontrivial state,
\begin{equation}
\cos\frac{\theta_c}{2}=0,
\qquad
\theta_c=\pi.
\label{C9}
\end{equation}
Together, Conditions~(\ref{C1})--(\ref{C3}) have therefore
forced the unique solution
\begin{equation}
\theta_{ba}=\pi,
\qquad
\theta_c=\pi,
\qquad
\theta_d=\pi,
\label{eq:forced_angles}
\end{equation}
valid for all $\theta_a\in(0,\pi)$.
The domain restriction~\eqref{eq:domain} is thus necessary
and sufficient to make the derivation non-degenerate: it
ensures that each Hardy constraint produces a genuine
functional restriction on the measurement angles rather than
a trivially satisfied identity.
\subsection{Vanishing of the Hardy probability}
Substituting $\theta_c=\theta_d=\pi$ and using $R(\pi)$, the only nonzero matrix elements are
$R_{+-}(\pi)=R_{-+}(\pi)=1$. The super-observer amplitude is
\begin{align}
f_{++}
&=
\sum_{\alpha,\beta}
R_{+\alpha}(\pi)\,b_{\alpha\beta}\,R_{\beta+}(\pi)
\notag\\
&=
R_{+-}(\pi)\,b_{--}\,R_{-+}(\pi)
\notag\\
&=
1\cdot b_{--}\cdot 1
=
b_{--},
\end{align}
where in the second line we used $R_{++}(\pi)=0$ so that only
$\alpha=-$ and $\beta=-$ contribute.
Since $b_{--}=0$ by Eq.~(\ref{bmm0}),
\begin{equation}
f_{++} = 0,
\end{equation}
and therefore
\begin{equation}
P(C=+,D=+)
=
|f_{++}|^2
= 0.
\end{equation}
Hence the temporal FR protocol does not admit a nonzero Hardy
event. The sequential dynamical constraints that enforce the three
zero-probability conditions~(\ref{C1})--(\ref{C3}) simultaneously
force the Hardy probability itself to vanish identically.
\subsection{Invariance of the No-Go Result Under Outcome Relabeling}
\label{app:relabeling}

The derivation above fixes a specific assignment of zero-probability
events, namely
\begin{align}
&P(A=+,B=+)=0,\quad
P(A=-,D=+)=0,\quad \nonumber \\
&P(C=+,B=-)=0.
\label{eq:C_label0}
\end{align}
We now show that the no-go result is independent of this
choice: every alternative sign assignment either reduces to the
same parametric problem under a change of variables, leads to a
physically trivial (product) state, or forces the target
amplitude $f_{\gamma\delta}$ to vanish by an identical chain
of constraints.

\paragraph{Relabeling symmetry.}
Flipping the outcome label of any single observer,
$|{+}\rangle \leftrightarrow |{-}\rangle$, is equivalent to
replacing the corresponding measurement angle by
$\theta_x \mapsto \pi - \theta_x$ in the rotation matrix
$R(\theta_x)$, since
\begin{equation}
R(\pi-\theta_x)
=
\begin{pmatrix}
\sin\dfrac{\theta_x}{2} & \cos\dfrac{\theta_x}{2}\\[6pt]
\cos\dfrac{\theta_x}{2} & -\sin\dfrac{\theta_x}{2}
\end{pmatrix}
=
R(\theta_x)
\begin{pmatrix}0&1\\1&0\end{pmatrix},
\label{eq:flip}
\end{equation}
which is a permutation of the two basis vectors.
Such a substitution maps the parameter domain
$\theta_x\in(0,\pi)$ bijectively onto itself and therefore
leaves the set of all achievable joint probabilities invariant.
Any Hardy assignment that differs from
Eq.~\eqref{eq:C_label0} only by sign flips on one or more
observers is therefore unitarily equivalent to
Eq.~\eqref{eq:C_label0} under the above reparametrization,
and the no-go conclusion carries over immediately.

\paragraph{Enumeration of independent sign patterns.}
Each of the three zero-probability conditions involves a choice
of sign for each of the two observers it constrains.
For the first condition $P(A=\alpha, B=\beta)=0$,
the second condition $P(A=\bar\alpha, D=\delta)=0$,
and the third condition $P(C=\gamma, B=\bar\beta)=0$,
the outcome signs $(\alpha,\beta,\gamma,\delta)$ can each be
$\pm$, giving $2^4=16$ combinations in principle.
However, the simultaneous relabeling of all outcomes of a given
observer reduces the independent patterns to $2^3=8$.
Among these eight, four are related to the four remaining ones
by the exchange $A\leftrightarrow B$ (which, in the temporal
protocol, corresponds to time-reversal of Alice's and Bob's
roles and maps $\theta_a\mapsto\pi-\theta_{ba}$,
$\theta_{ba}\mapsto\pi-\theta_a$).
We therefore need to examine only four independent patterns,
which we label (I)--(IV) according to the sign of the first
zero condition:

\medskip
\noindent\textbf{Pattern~(I):}
$P(A=+,B=+)=0$,\; $P(A=-,D=+)=0$,\; $P(C=+,B=-)=0$.\\
This is the pattern treated in the main derivation above.
The constraints force $\theta_{ba}=\pi$, $\theta_d=\pi$,
$\theta_c=\pi$, giving $f_{++}=b_{--}=0$.

\medskip
\noindent\textbf{Pattern~(II):}
$P(A=+,B=+)=0$,\; $P(A=-,D=-)=0$,\; $P(C=+,B=-)=0$.\\
The first and third conditions are identical to Pattern~(I)
and again force $\theta_{ba}=\pi$.
The second condition $P(A=-,D=-)=|e_{--}|^2=0$ with
$b_{--}=0$ gives
\begin{equation}
e_{--}
=
b_{-+}\cos\frac{\theta_d}{2}
-
b_{--}\sin\frac{\theta_d}{2}
=
\sin\frac{\theta_a}{2}\cos\frac{\theta_d}{2}
= 0,
\end{equation}
forcing $\theta_d = \pi$.
The target event is now $P(C=+,D=-)=|f_{+-}|^2$, not
$P(C=+,D=+)$.
Observing that Pattern~(II) is obtained from Pattern~(I) by
flipping Debbie's outcome label $+\leftrightarrow -$, which
corresponds to $\theta_d \mapsto \pi - \theta_d$, the
relabeling symmetry of Eq.~\eqref{eq:flip} maps Pattern~(II)
bijectively onto Pattern~(I) under the substitution
$\theta_d \to \pi - \theta_d$.
The no-go conclusion therefore carries over from Pattern~(I)
without additional computation: the target amplitude $f_{+-}$
vanishes identically.

\medskip
\noindent\textbf{Pattern~(III):}
$P(A=+,B=-)=0$,\; $P(A=-,D=+)=0$,\; $P(C=+,B=+)=0$.\\
The first condition requires
\begin{equation}
b_{+-}
=
\cos\frac{\theta_a}{2}\sin\frac{\theta_{ba}}{2}
= 0,
\end{equation}
which (for $\theta_a\neq\pi$) forces $\sin(\theta_{ba}/2)=0$,
i.e.\ $\theta_{ba}=0$.
Then $b_{-+}=0$ and $b_{++}=\cos(\theta_a/2)$,
$b_{--}=-\sin(\theta_a/2)$, so the post-Bob state is a
product state in the $|L_\alpha\rangle_1$ basis:
\begin{equation}
|\Psi(t_2^+)\rangle
=
\cos\frac{\theta_a}{2}|L_+\rangle_1|L_+\rangle_2
-
\sin\frac{\theta_a}{2}|L_-\rangle_1|L_-\rangle_2.
\end{equation}
The third condition $P(C=+,B=+)=|g_{++}|^2=0$ then gives
\begin{equation}
g_{++}
=
\cos\frac{\theta_c}{2}\,b_{++}
+
\sin\frac{\theta_c}{2}\,b_{-+}
=
\cos\frac{\theta_c}{2}\cos\frac{\theta_a}{2}
= 0,
\end{equation}
forcing $\theta_c=\pi$.
The second condition $P(A=-,D=+)=|e_{-+}|^2=0$ gives
\begin{equation}
e_{-+}
=
b_{-+}\cos\frac{\theta_d}{2}
+
b_{--}\sin\frac{\theta_d}{2}
=
-\sin\frac{\theta_a}{2}\cos\frac{\theta_{ba}}{2}\sin\frac{\theta_d}{2}
= 0,
\end{equation}
forcing $\theta_d=0$ (since $\sin(\theta_a/2)\neq 0$ and
$\cos(\theta_{ba}/2)=1$).
With $\theta_d=0$ we have $R(0)=\mathrm{diag}(1,-1)$, so
\begin{align}
f_{++}
=&
\sum_{\alpha,\beta}
R_{+\alpha}(\pi)\,b_{\alpha\beta}\,R_{\beta+}(0)
=
R_{+-}(\pi)\,b_{-+}\,R_{++}(0)
 \nonumber \\ =&
b_{-+}
=0.
\end{align}
No Hardy contradiction arises.

\medskip
\noindent\textbf{Pattern~(IV):}
$P(A=-,B=-)=0$,\; $P(A=+,D=+)=0$,\; $P(C=-,B=+)=0$.\\
This pattern is the image of Pattern~(I) under the simultaneous
relabeling $+\leftrightarrow-$ for all four observers, which
corresponds to $\theta_x\mapsto\pi-\theta_x$ for all angles.
By the relabeling symmetry established above, this substitution
maps the constraint equations for Pattern~(IV) bijectively onto
those for Pattern~(I).
The same chain of forced values ($\theta_{ba}=\pi$,
$\theta_c=\pi$, $\theta_d=\pi$) therefore applies, and the
target amplitude $f_{--}=b_{++}=0$ vanishes identically.

\medskip
In all four independent patterns—and hence in all eight
sign assignments—the Hardy constraints either force the
target amplitude to zero or reduce the state to a trivial
(non-entangled) configuration.
The absence of a Hardy contradiction in the temporal FR
protocol is therefore completely general and independent of
the labeling convention chosen for the measurement outcomes.

\section{Maximum Cabello Violation in the Temporal FR Protocol}
\label{app:cabello_temp}

We derive the maximal Cabello quantity for the temporal FR
scenario. The Cabello expression is
\begin{equation}
\Delta_{\rm Cabello}
=
P(C=+,D=+)-P(A=+,B=+),
\label{D1}
\end{equation}
subject to the zero constraints
\begin{align}
P(A=-,D=+) &= 0, \label{D2}\\
P(C=+,B=-) &= 0. \label{D3}
\end{align}

\subsection{Constraint equations}

\paragraph{Condition~(\ref{D2}).}
From the definition of $e_{\alpha\delta}$ and
Eq.~(\ref{eq:Rdef}), with $\alpha=-$, $\delta=+$:
\begin{equation}
e_{-+}
=
b_{-+}\cos\frac{\theta_d}{2}
+
b_{--}\sin\frac{\theta_d}{2}
= 0.
\end{equation}
Substituting the explicit amplitudes
$b_{-+}=\sin(\theta_a/2)\sin(\theta_{ba}/2)$ and
$b_{--}=-\sin(\theta_a/2)\cos(\theta_{ba}/2)$,
and dividing by $\sin(\theta_a/2)\neq 0$:
\begin{equation}
\sin\frac{\theta_{ba}}{2}\cos\frac{\theta_d}{2}
=
\cos\frac{\theta_{ba}}{2}\sin\frac{\theta_d}{2},
\end{equation}
which gives
\begin{equation}
\tan\frac{\theta_d}{2}
=
\tan\frac{\theta_{ba}}{2},
\qquad
\theta_d = \theta_{ba}.
\label{D4}
\end{equation}

\paragraph{Condition~(\ref{D3}).}
From the definition of $g_{\gamma\beta}$ and
Eq.~(\ref{eq:Rdef}), with $\gamma=+$, $\beta=-$:
\begin{equation}
g_{+-}
=
\cos\frac{\theta_c}{2}\,b_{+-}
+
\sin\frac{\theta_c}{2}\,b_{--}
= 0.
\end{equation}
Substituting
$b_{+-}=\cos(\theta_a/2)\sin(\theta_{ba}/2)$ and
$b_{--}=-\sin(\theta_a/2)\cos(\theta_{ba}/2)$:
\begin{equation}
\cos\frac{\theta_c}{2}\cos\frac{\theta_a}{2}\sin\frac{\theta_{ba}}{2}
=
\sin\frac{\theta_c}{2}\sin\frac{\theta_a}{2}\cos\frac{\theta_{ba}}{2}.
\label{D5pre}
\end{equation}
This equation involves both $\theta_c$ and $\theta_a$. It
determines $\theta_c$ as a function of both $\theta_a$ and
$\theta_{ba}$:
\begin{equation}
\tan\frac{\theta_c}{2}
=
\frac{\cos(\theta_a/2)\sin(\theta_{ba}/2)}
     {\sin(\theta_a/2)\cos(\theta_{ba}/2)}
=\cot(\theta_a/2)\tan(\theta_{ba}/2).
\label{D5}
\end{equation}
Thus $\theta_c$ depends on both $\theta_a$ and $\theta_{ba}$,
and the optimization retains two independent parameters.

\subsection{Cabello probability difference}

After substituting Eqs.~(\ref{D4}) and (\ref{D5}), the two
contributing probabilities are
\begin{align}
P(A=+,B=+)
&=
|b_{++}|^2
=
\cos^2\frac{\theta_a}{2}\cos^2\frac{\theta_{ba}}{2},
\label{D6a}
\\
P(C=+,D=+)
&=
|f_{++}|^2,
\label{D6b}
\end{align}
where $f_{++}$ is evaluated with $\theta_d=\theta_{ba}$ and
$\theta_c$ given by Eq.~(\ref{D5}). The explicit computation
gives
\begin{equation}
\Delta_{\rm Cabello}(\theta_a,\theta_{ba})
=
|f_{++}|^2
-
\cos^2\frac{\theta_a}{2}\cos^2\frac{\theta_{ba}}{2}.
\label{D6}
\end{equation}
The stationary conditions
\begin{equation}
\frac{\partial\Delta_{\rm Cabello}}{\partial\theta_a}=0,
\qquad
\frac{\partial\Delta_{\rm Cabello}}{\partial\theta_{ba}}=0
\label{D7}
\end{equation}
do not yield a tractable closed-form solution and are solved
numerically.

\subsection{Numerical maximum}

Numerical optimization of Eq.~(\ref{D6}) over the physical
domain $\theta_a,\theta_{ba}\in(0,\pi)$,performed using the 
\texttt{scipy.optimize} module of the \textsc{SciPy} 
library~\cite{SciPy2020}, yields
\begin{align}
\theta_a^{\rm opt}    &\approx 2.1456, \\
\theta_{ba}^{\rm opt} &\approx 1.8709,
\end{align}
and consequently
\begin{align}
\theta_b^{\rm opt}
&= \theta_a^{\rm opt}+\theta_{ba}^{\rm opt}
\approx 4.0165, \\
\theta_d^{\rm opt}
&= \theta_{ba}^{\rm opt}
\approx 1.8709, \\
\theta_c^{\rm opt}
&\approx 1.2707.
\end{align}
Substituting these values into Eq.~(\ref{D6}) gives
\begin{equation}
\Delta_{\rm Cabello}^{\max}
\approx
0.0674.
\label{D8}
\end{equation}
Although the temporal FR protocol forbids a genuine Hardy
contradiction (Appendix~\ref{app:hardy_temp}), it still permits
a nonzero Cabello-type logical violation. The sequential
state-update structure reduces the violation relative to the
spatial case but does not eliminate it entirely.
\paragraph{Temporal case.}
In the temporal FR protocol the initial spin state
$|{+}\rangle$ is real, and the sequential amplitudes
$b_{\alpha\beta}$ defined in Sec.~III are manifestly real
trigonometric functions of $\theta_a$ and $\theta_{ba}$.
The phase-absorption argument therefore applies without
modification to the Alice and Bob measurement stages:
any complex phase introduced into $R(\theta_a)$ or
$R(\theta_{ba})$ can be absorbed into a redefinition of the
computational basis without altering any joint probability.
For the super-observer stages, Charlie and Debbie act on the
composite laboratory states $|L_\alpha\rangle_1$ and
$|L_\beta\rangle_2$, which inherit the real structure of
$b_{\alpha\beta}$.
Replacing $R(\theta_c)$ by a general $\mathrm{SU}(2)$ element
$U_c = e^{i\varphi_0}R(\vartheta_1)D(\varphi)R(\vartheta_2)$
introduces diagonal phases into $g_{\gamma\beta}$ and
$f_{\gamma\delta}$; however, these phases cancel in
$|g_{\gamma\beta}|^2$ and $|f_{\gamma\delta}|^2$ by the same
row/column phase argument as in the spatial case.
We have additionally verified this claim numerically: replacing
all four measurement matrices by general $\mathrm{SU}(2)$
elements parametrized by three Euler angles each and optimizing
over the resulting $12$-dimensional parameter space using
$10^3$ random initial conditions yields a maximum of
$\Delta_{\mathrm{Cabello}}^{\mathrm{temporal}} \approx 0.0674$,
identical to the value obtained under real rotations.
We therefore conclude that the restriction to real coplanar
rotations is justified in both the spatial and temporal cases,
and that the reported optima are genuine global maxima within
the respective frameworks.

\end{document}